\documentclass{article}
\usepackage[utf8]{inputenc}
\usepackage{graphicx}
\usepackage{wrapfig}
\usepackage{lipsum}
\usepackage{physics}
\usepackage{caption}
\usepackage{subcaption}
\usepackage{setspace}
\usepackage{authblk}
\usepackage[hyphens]{url}
\usepackage{hyperref}
\usepackage[nottoc]{tocbibind}
\usepackage{float}
\usepackage{mathtools}
\usepackage{subcaption}
\usepackage{color}
\usepackage[dvipsnames]{xcolor}
\usepackage{breqn}
\usepackage{url}
\usepackage{verbatim}
\usepackage{amsmath}
\usepackage{multicol}
\usepackage[margin=0.9in]{geometry}
\usepackage[labelfont=bf,labelsep=space,font=footnotesize]{caption}

\title{\textbf{\Large{Systematic improvement of molecular excited state calculations by inclusion of nuclear quantum motion: \\a mode-resolved picture and the effect of molecular size}}}

\author[1]{Timothy J. H. Hele}
\author[2,3]{Bartomeu Monserrat}
\author[3,*]{Antonios M. Alvertis}

\affil[1]{Department of Chemistry, University College London, 20, Gordon Street, London WC1H 0AJ, United Kingdom}
\affil[2]{Department of Materials Science and Metallurgy, University of Cambridge, 27 Charles Babbage Road, Cambridge CB3 0FS, United Kingdom}
\affil[3]{Cavendish Laboratory, University of Cambridge, J.\,J.\,Thomson Avenue, Cambridge CB3 0HE, United Kingdom}
\affil[*]{e-mail: ama80@cam.ac.uk}


\date{}

\begin{document}

\maketitle

\noindent
\section*{Abstract}

\noindent
The energies of molecular excited states arise as
solutions to the electronic Schr\"{o}dinger equation and are often compared
to experiment. At the same time, nuclear quantum motion is known to be important and to induce a 
red-shift of excited state energies.
However, it is thus far unclear whether incorporating nuclear quantum motion in molecular excited state calculations leads to a systematic
improvement of their predictive accuracy, making further investigation
necessary.
Here we present such an investigation by employing two first-principles
methods for capturing the effect of quantum fluctuations on excited state energies, which we apply to the
Thiel set of organic molecules. We show that accounting
for zero-point motion leads to
much improved agreement with experiment, compared to `static' calculations
which only account for electronic effects, and the magnitude of the red-shift
can become as large as $1.36$\,eV. Moreover, we show that the effect of nuclear quantum motion on excited state
energies largely depends on the molecular size, with smaller molecules
exhibiting larger red-shifts. Our methodology also makes it possible to
analyze the contribution of individual vibrational normal modes to the
red-shift of excited state energies, and in several molecules we identify
a limited number of modes dominating this effect. 
Overall, our study provides
a foundation for systematically quantifying the shift
of excited state energies due to nuclear quantum motion, and for understanding this effect at a microscopic level.

\clearpage

\section{Introduction}
\label{introduction}
The optoelectronic properties of organic molecules are dominated by their low-lying excited states known as excitons \cite{Frenkel1931}. Exciton energies are 
critical to several technologically-relevant processes in these systems,
such as singlet fission and thermally activated delayed fluorescence,
which find applications in photovoltaics and LEDs respectively 
\cite{Rao2017,Alvertis2019_2,Reineke2014,Evans2018}. It is therefore desirable to develop methods for the accurate 
prediction of exciton energies. 

Typical methods for the calculation of excited state energies, such
as time-dependent density functional theory (TD-DFT) \cite{Petersilka1996}, coupled cluster (CC) \cite{Bartlett1989}, complete active space self-consistent ﬁeld (CAS-SCF) \cite{Malmqvist1989}, and the Bethe-Salpeter equation (BSE) 
\cite{Rohlfing2000,Jacquemin2015,Jacquemin2017}, only account for electronic contributions to exciton states, typically computing vertical excitation energies at a fixed geometry of the system. However, in a
recent study, Bai \emph{et al.} \cite{Bai2020} showed that the vibrational motion of molecules is
responsible for a red-shift of the absorption maximum compared to `static' vertical
excitation energies, an effect which needs to be accounted for in order to achieve
predictive accuracy. Even
at $0$\,K, atomic nuclei vibrate with a zero-point energy of
$\frac{1}{2}\hbar \omega$ per normal mode. In a recent study on solid state 
organic semiconductors \cite{Alvertis2020}, two of us showed that
this nuclear quantum motion can significantly change the `static' exciton energies that are commonly
computed at the
ground state geometry of a system, and that incorporating these effects leads to improved agreement with
experiment.

A number of computational studies have proposed advanced methods for accurately 
simulating the shape of molecular absorption spectra including vibrational effects
\cite{Santoro2007,Charaf-Eddin2014,Baiardi2013}. Additionally,
the effect of nuclear quantum fluctuations on molecular excited states is now understood
to cause a red-shift of exciton energies compared to their `static' values \cite{Bai2020}.
Despite these advances, there remain a number of open questions regarding
the effect of quantum fluctuations on molecular excited states. In particular: 
(i) It remains unclear to what extent the inclusion of quantum fluctuations leads to a systematic improvement of molecular excited state calculations, in terms of agreement with experiment for the exciton energy, which corresponds to the frequency of the absorption maximum in the case
of a state with non-zero oscillator strength.
(ii) How many vibrational normal modes significantly contribute to the renormalization of exciton
energies due to quantum motion is, to the best of our knowledge, not yet understood,
since the studies which have emerged so far do not provide a mode-resolved
picture of this effect. (iii) Our previous work in periodic structures \cite{Alvertis2020} suggests that 
the change in the exciton energy due to zero-point motion depends on the size of the studied system. It is hence 
important to investigate
whether such a trend also holds for isolated organic molecules. 
(iv) It is thus far not clear whether the level
of theory employed for the calculation of the zero-point renormalization of the exciton energy has a large impact on its value, or whether the magnitude of this effect is
largely independent of the underlying level of calculations. This last point is particularly important for practical calculations, as such an independence would imply
that one could compute the correction due to nuclear quantum effects at a cheaper level of theory than the `static' exciton energy, hence reducing the overall computational cost of the calculation.

In this work we systematically investigate the effect of nuclear quantum motion on exciton energies of organic molecules and address the aforementioned questions. In order
to accurately compute the zero-point renormalization of exciton energies, we employ
a Monte Carlo sampling technique, combining TD-DFT and finite difference
methods for the molecular vibrations \cite{Kunc1982,Monserrat2018}. This Monte Carlo method
had thus far been used in the context of periodic systems \cite{Alvertis2020} and allows us to capture exciton-vibration interactions to all orders and to treat
excited state energy surfaces without any harmonic assumption. It also has 
strong similarities to the nuclear ensemble approach \cite{Crespo-Otero2012}, which
has been developed for molecular systems and is also used in Ref. \cite{Bai2020}. 

While both the nuclear ensemble and Monte Carlo methods capture the renormalization
of exciton energies due to quantum fluctuations, they do not provide information on the
individual contribution of normal modes to this effect. For example, low- and high-frequency vibrations in organic
systems are known to play different roles during vibrational relaxation \cite{Alvertis2019} and electron-transfer reactions \cite{Rafiq2021}, 
therefore obtaining a mode-resolved picture
for the exciton energy renormalization is critical to obtaining microscopic insights into the physics of
these systems.
Such mode-resolved information is available
through the use of a quadratic approximation to the exciton-vibration coupling \cite{Monserrat2018}, however this comes at the cost of only capturing these interactions
to third order. Here we employ the quadratic approximation for capturing the correction to exciton energies induced by nuclear quantum motion, and we assess the accuracy
of this method by comparing to the more accurate Monte Carlo calculations.

\begin{figure}[tb]
\centering
\includegraphics[width=0.8\linewidth]{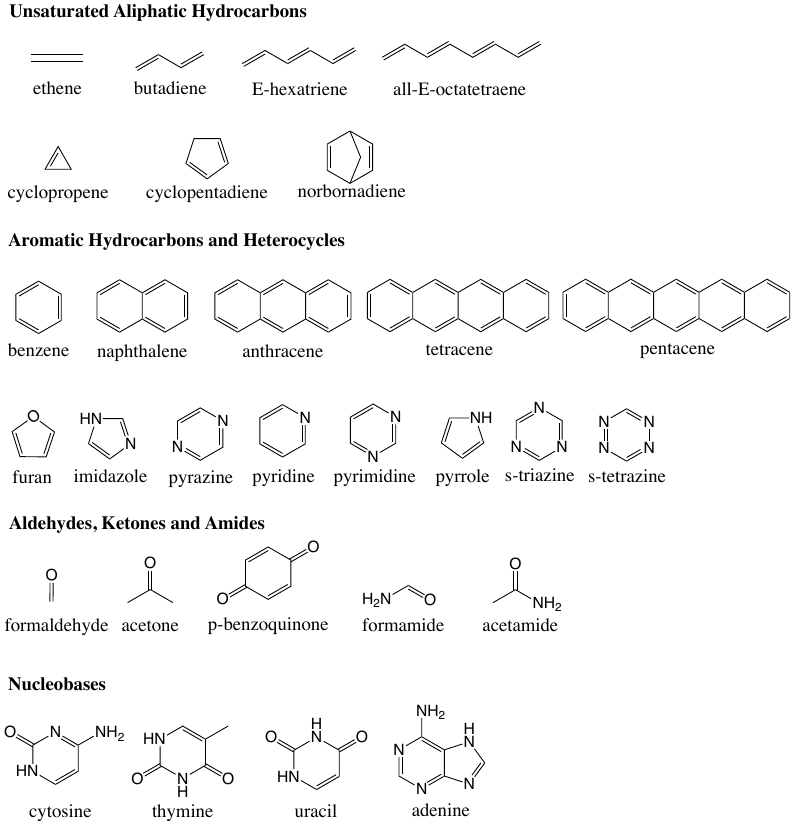}
\caption{Molecular structures studied in this work.}
\label{fig:structures}
\end{figure}

We apply our set of methods to the 
so-called Thiel set of organic molecules (sometimes also referred to as the M\"{u}lheim set), for which highly accurate exciton energies have been computed using
wavefunction-based methods \cite{Schreiber2008}. The Thiel set of molecules consists
of four broad categories of structures: unsaturated aliphatic hydrocarbons, aromatic hydrocarbons and heterocycles, carbonyl compounds, and nucleobases. The
studied structures are shown in Figure\,\ref{fig:structures}. In addition to the molecules that are commonly included in the
Thiel set, we also study three additional molecules: anthracene, tetracene and pentacene, which together with benzene and naphthalene that are included in the Thiel set form the first five members of the acene family. We have also excluded propanamide and pyridazine from the studied structures, as we were unable to obtain converged vibrational properties for these molecules.
It has become common in the literature to compare computational methods for the calculation of exciton 
energies to the Thiel set 
\cite{Roychoudhury2020,Jacquemin2015,Bruneval2015,Rangel2017,Jacquemin2017,Bai2020}, a path
that we also take. Moreover, throughout the paper we systematically compare our results to experiment and find that both the Monte Carlo and quadratic methods for computing
the renormalization of the exciton energies due to zero-point motion of the nuclei are remarkably accurate.
This allows us to draw several important conclusions on the microscopic mechanism of
this phenomenon and to assess different methods of describing it accurately. 

The structure of this paper is as follows. In section \ref{background} we provide the necessary
theoretical background for the results that
follow in section \ref{Results}. Specifically, in subsection \ref{basic_background} we include a qualitative discussion of the effect of molecular
vibrations on exciton energies and based on standard expressions we lay the foundations to discuss
the Monte Carlo sampling technique and the quadratic approximation for
quantifying this effect, in subsections \ref{Monte_Carlo} and \ref{quadratic} respectively. In the results section \ref{Results} we
compare our results from a Monte
Carlo sampling of the exciton energies to experiment and to previous 
computational studies in subsection \ref{comparisons}. The effect of using different levels of electronic structure theory on the predicted magnitude of the exciton energy renormalization is examined in subsection \ref{levels_of_theory}, and the impact of the molecular size on the exciton energy correction is discussed in subsection \ref{molecular_size}. We then consider the accuracy and speed
of the quadratic method in subsection \ref{quadratic_accuracy}, and proceed to use it in order to attain a mode-resolved picture of the effect
of nuclear quantum motion on exciton energies in subsection \ref{mode_resolved}. Finally, we summarize our results and conclude our study in section \ref{conclusions}.

\section{Theoretical background and computational methods}
\label{background}

\subsection{Effect of vibrations on exciton energies}
\label{basic_background}
\begin{figure}[tb]
\centering
\includegraphics[width=\linewidth]{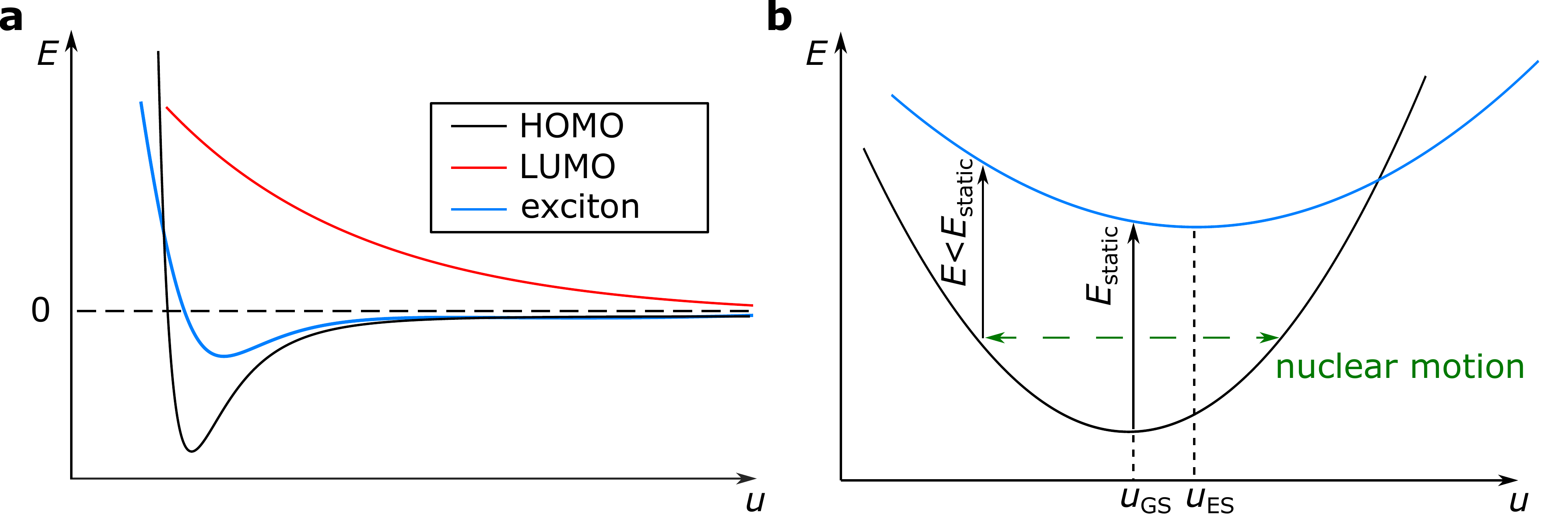}
\caption{Schematic representation of the energy of the highest occupied (HOMO) and lowest unoccupied (LUMO) molecular orbitals,
as well as the exciton state that results from a transition between them,
along a generalized coordinate $u$ (panel \textbf{a}). Due to the excited
state potential energy surface having a lower curvature, nuclear
motion leads to a red-shift of the energy difference between the ground
and excited states compared to its static value (panel \textbf{b}).}
\label{fig:schematic}
\end{figure}

We start by presenting standard results and a qualitative discussion of the effects of molecular vibrations on exciton energies, which help inform
our later discussion.
In organic molecules, the highest occupied molecular orbital (HOMO) is a bonding $\sigma$ or $\pi$ orbital that lowers the energy
of the molecule once occupied and leads to nuclei that are closer to each other. Therefore, the energy of such an orbital along a generalized nuclear
coordinate $u$ of the molecular system will look similar to a 
Morse potential, as visualized in Figure\,\ref{fig:schematic}a 
(black). The lowest unoccupied molecular orbital (LUMO) is an anti-bonding $\sigma^*$ or $\pi^*$ orbital that raises the energy of the molecule once occupied and forces
the nuclei apart, its energy along $u$ showing an exponential
decay (Figure\,\ref{fig:schematic}a, red). For most conjugated organic molecules,
the lowest energy singlet exciton ($\text{S}_1$) is formed by
exciting an electron from the HOMO to the LUMO \cite{AtkinsP.W.PeterWilliam2005Mqm}, with energy:

\begin{align}
    \label{eq:Coulomb_singlet}
    E(\text{S}_1)=E(\text{S}_0)+\epsilon_L-\epsilon_H-J_{HL}+2K_{HL}.
\end{align}

\noindent
Here $E(\text{S}_0)$ is the energy of the ground state,  $\epsilon_H$ and $\epsilon_L$ are the energies of the HOMO and the LUMO respectively, $J_{HL}$ the HOMO-LUMO Coulomb
integral and $K_{HL}$ the HOMO-LUMO
exchange integral. 
By using
equation\,\ref{eq:Coulomb_singlet}, we schematically describe the energy
of an exciton along $u$ (Figure\,\ref{fig:schematic}a, blue), having 
assumed that the dependence of the integrals $J,K$ on $u$ does not alter its qualitative 
characteristics. 

What can be observed from Figure\,\ref{fig:schematic}a is that
the excited state potential energy curve (blue) has a lower curvature compared to the ground state one (black) due to the contribution of the anti-bonding orbital. Therefore, if we approximate the two within the harmonic
approximation, we obtain two parabolas of different curvature, as shown
in Figure \ref{fig:schematic}b. If the structure was `frozen' at its
ground state configuration $u_{\text{GS}}$, then the energy required
to access the exciton state would be $E_{\text{static}}$, which
is the vertical excitation energy usually computed by electronic structure calculations. However, due to the vibrational motion of the nuclei, the system can explore
a distribution of configurations within the region denoted by green
arrows, even at $0$\,K. If we excited the system from any given displaced configuration
within the available region, then the excitation energy becomes
smaller than $E_{\text{static}}$, since the lower parabola 
corresponding to the ground state has a higher curvature than the excited state
one. Therefore, it is expected that inclusion of the effects
of nuclear fluctuations will lead to a red-shift of exciton
energies. In particular, at temperature $T$ the absorption maximum will not be found at the
energy $E_{\text{static}}$ corresponding to vertical excitation, but at
a lower energy, corresponding to the quantum mechanical vibrational average :

\begin{equation}
    E_{\text{exc}}(T)=\frac{1}{\mathcal{Z}}\sum_{\mathbf{s}}\bra{\chi_{\mathbf{s}}(\boldsymbol{u})}E_{\text{exc}}(\boldsymbol{u})\ket{\chi_{\mathbf{s}}(\boldsymbol{u})} e^{-E_{\mathbf{s}}/k_{\mathrm{B}}T}, \label{eq:exp_value}
\end{equation}

\noindent
where $\ket{\chi_{\mathbf{s}}(\boldsymbol{u})}$ is a vibrational eigenstate on
the ground state potential energy surface with energy  $E_{\mathbf{s}}$, 
$\mathcal{Z}=\sum_{\mathbf{s}}e^{-E_{\mathbf{s}}/k_{\mathrm{B}}T}$ is 
the partition function, and $\boldsymbol{u}$ is the nuclear displacement.

An intuitive way of representing this mean exciton energy on a diagram of the
potential energy surfaces of a molecule, is to plot the transition energy
at a `mean value' configuration $\boldsymbol{u}^{MV}$ where the vertical exciton energy is equal to
the average at temperature $T$ \cite{Monserrat2016}:

\begin{equation}
    E_{\text{exc}}(\boldsymbol{u}^{MV})=E_{\text{exc}}(T).
\end{equation}

\begin{figure}[tb]
\centering
\includegraphics[width=\linewidth]{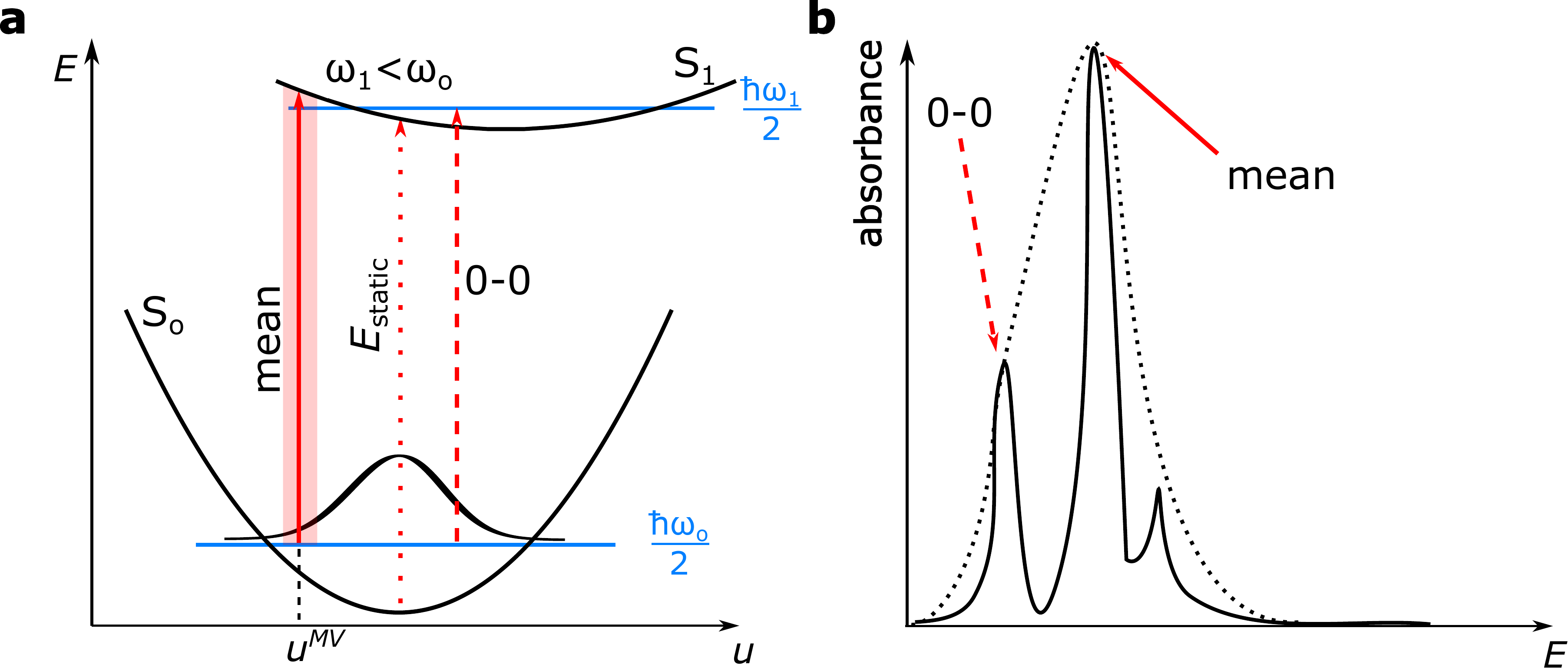}
\caption{Vibrational effects result in a distribution of exciton energies (panel \textbf{a}, red shaded region) around a mean value, which is different to the `static' exciton energy $E_{\text{static}}$. This can be
represented as the vertical transition from the zero-point level of the ground state, to the excited state surface, at a `mean value' configuration $u^{MV}$ (panel \textbf{a}).  This mean
energy of the excited state appears as
the band maximum in experimental absorption spectra (panel \textbf{b}). In contrast to that,
the $0-0$ energy refers to the energy difference between the zero-point
levels of the optimized ground and excited states. These separate peaks
cannot always be resolved experimentally, in which cases an envelope function (panel \textbf{b}, dotted lined) is found instead.}
\label{fig:max_vs_0-0}
\end{figure}

\noindent
According to the mean-value theorem for integrals, there always exists such
a configuration $u^{MV}$. We can therefore visualize the mean exciton energy,
corresponding to the absorption band maximum in Figure\,\ref{fig:max_vs_0-0}.
It is also worth noting that the absorption maximum described by the
quantum mechanical expectation value of equation\,\ref{eq:exp_value} is
distinct from the so-called $0-0$ energy that is often reported in the
literature, and which corresponds to the energy difference between the zero-point
levels of the optimized ground and excited state configurations, as visualized in Figure\,\ref{fig:max_vs_0-0}.

To simplify the problem of computing the exciton band maximum energy from equation\,\ref{eq:exp_value}, we use the harmonic approximation
for the ground state potential, and by substituting $\ket{\chi_{\mathbf{s}}(\boldsymbol{u})}$ with the wavefunction of a quantum harmonic
oscillator we obtain \cite{Monserrat2018}:

\begin{equation}
    E_{\text{exc}}(T)=\int d\boldsymbol{u}|\Phi(\boldsymbol{u};T)|^2E_{\text{exc}}(\boldsymbol{u}), \label{eq:harm_exp_value}
\end{equation}

\noindent
where $E_{\text{exc}}(\boldsymbol{u})$ is the vertical excitation energy
at the configuration $\boldsymbol{u}$ and:

\begin{equation}
    |\Phi(\boldsymbol{u};T)|^2 = \prod_{\nu}(2\pi \sigma^2_{\nu}(T))^{-1/2}\exp{\left(-\frac{u_{\nu}^2}{2\sigma^2_{\nu}(T)}\right)},
    \label{eq:harm_density}
\end{equation}

\noindent
is the harmonic density at temperature $T$, which in turn is a product of Gaussian functions of width:

\begin{equation}
    \sigma^2_{\nu}(T) = \frac{1}{2\omega_{G\nu}}\cdot \coth{\left(\frac{\omega_{G\nu}}{2k_BT}\right)}.
    \label{eq:Gaussian_width}
\end{equation}

\noindent
In the above, atomic units and mass-weighted coordinates have 
been used, and $\nu$ is the index labeling the ground state 
vibrational modes of the studied molecule, with $\omega_{G\nu}$ the
corresponding frequency. For non-linear molecular systems there are six vibrational
modes with zero frequency which correspond to translations and
rotations of the entire structure. These are not included in the sampling of the integral
of equation\,\ref{eq:harm_exp_value}.

\subsection{Monte Carlo sampling of the exciton energy}
\label{Monte_Carlo}

The Monte Carlo method approximates the integral in equation\,\ref{eq:harm_exp_value} by 
generating configurations $\boldsymbol{u}$ that are distributed according to the
harmonic density at the temperature of choice $T$. Then, the expectation value of 
equation\,\ref{eq:harm_exp_value} is computed as the simple average of
the computed values of the vertical exciton energy $E_{\text{exc}}(\boldsymbol{u})$ 
at the displaced configurations. This methodology for capturing the effect
of vibrations on exciton energies has previously been applied in the
context of periodic organic crystals \cite{Alvertis2020}, and here we
extend it to isolated organic molecules. This computational approach has the advantage that it
makes no assumption about the shape of the excited state potential
energy surface and it also includes exciton-vibration interactions to all
orders. Moreover, it
relies on no
adjustable parameters, apart from those in the DFT functional, which are fixed throughout the entire series of calculations. This Monte Carlo
approach is conceptually very similar to the nuclear ensemble method
\cite{Crespo-Otero2012}, which was recently applied to the Thiel set
of molecules \cite{Bai2020}, and we compare our results to this previous study later on, in Figure\,\ref{fig:comparison_Barbatti}.
For all the molecules studied we
generate $100$ displaced configurations at $T=0$\,K and $T=300$\,K and
their exciton energies are computed through 
TD-DFT within the Tamm-Dancoff approximation \cite{Hirata1999}, using the popular B3LYP hybrid functional \cite{Becke1993} and
the cc-pVDZ basis set, as implemented within the NWChem code \cite{Apra2020}. 
We hence obtain renormalized exciton energies compared to those of 
a `static' TD-DFT calculation for the Thiel set of molecules. We generally find that approximately $50$ points are sufficient to sample
the expectation value of the exciton energies (SI section\,S4). The numerical results of these calculations (and the associated statistical uncertainties) are given in SI section\,S1, not only at the B3LYP level, but also for two additional DFT functionals as discussed in subsection \ref{levels_of_theory}.

Since the Monte Carlo method provides an approximation to the expectation
value of equation\,\ref{eq:exp_value}, we always compare the computed
energy values to the absorption
maximum which corresponds to the same exciton state (for example the first bright excited state), 
and not to the so-called $0-0$
energy (see also Figure\,\ref{fig:max_vs_0-0}). 
However, it is not uncommon that experimental works report the energy of the $0-0$ transition.
For the vast majority of systems studied here, and unlike  the schematic of Figure\,\ref{fig:max_vs_0-0} which aims to emphasize the
differences between these two quantities, the energies of the $0-0$ transition and the band maximum are either identical or very close to each other. 
Accurate explicit calculations of
$0-0$ energies have been reported in the literature \cite{Loos2018}, 
and these require finding the optimized geometry of the excited state
and computing its Hessian matrix, which can be very challenging computationally, especially for larger molecules such as the acene series
and nucleobases, which are studied here. Given that the energy of the $0-0$
transition is in most cases very close, or even identical to that of the
band maximum, the Monte Carlo
method (and also the quadratic approximation which is outlined in the
following subsection) could potentially be used to also approximate $0-0$ energies at a much lower
computational cost.

\subsection{The quadratic approximation}
\label{quadratic}

One can further simplify the expression of equation\,\ref{eq:harm_exp_value} by performing a quadratic expansion of
the vertical excitation energy  $E_{\text{exc}}(\boldsymbol{u})$ in the
coordinates $u_{\nu}$ of the vibrational normal modes, yielding:

\begin{equation}
    \label{eq:quadtatic_expansion}
    E_{\text{exc}}(\boldsymbol{u}) = E_{\text{exc}}(\boldsymbol{0})+\sum_{\nu} \frac{\partial E_{\text{exc}}(\boldsymbol{0})}{\partial u_{\nu}}u_{\nu}+\frac{1}{2}\sum_{\nu}\sum_{\nu'}\frac{\partial^2 E_{\text{exc}}(\boldsymbol{0})}{\partial u_{\nu}\partial u_{\nu'}}u_{\nu}u_{\nu'}+... \hspace{0.1cm}.
\end{equation}

\noindent
Here $E_{\text{exc}}(\boldsymbol{0})$ is the exciton energy at the ground
state geometry and is equal to the static exciton energy $E_{\text{static}}$ as defined previously. 
Substituting this expression in equation\,\ref{eq:harm_exp_value} gives:

\begin{equation}
    \label{eq:quadratic}
    E_{\text{exc}}(T) = E_{\text{static}}+\sum_{\nu} \frac{1}{2\omega_{\nu}}\cdot \frac{\partial^2E_{\text{exc}}}{\partial u_{\nu}^2}[\frac{1}{2}+n_B(\omega_{\nu},T)]+\mathcal{O}(u^4).
\end{equation}

\noindent
In equation\,\ref{eq:quadratic}, $n_B(\omega_{\nu},T)$ is the Bose-Einstein
distribution for the vibrational quanta at temperature $T$. When computing
the expectation value of the exciton energy in 
equation\,\ref{eq:quadratic}, all odd terms vanish due to the harmonic
density $|\Phi(\boldsymbol{u};T)|^2$ being an even function, thus the
resulting approximation is accurate to fourth order in $u$, and the
exciton energy renormalization is described to lowest order by the
quadratic term of the expansion. While this so-called quadratic 
approximation is in principle less accurate than the Monte Carlo approach
which we outlined previously, it has the advantage of separating the
contributions of the different vibrational normal modes to the exciton 
energy renormalization, potentially allowing for additional microscopic
insights into these effects. 

Equation\,\ref{eq:quadratic} is used for
calculations with the quadratic method, which essentially involves the calculation of the second derivative of the exciton
energy along each vibrational mode of the system by using the finite
difference formula:

\begin{equation}
    \label{eq:second_der}
    \frac{\partial^2E_{\text{exc}}}{\partial u_{\nu}^2} \approx \frac{E_{\text{exc}}(\delta u_{\nu})+E_{\text{exc}}(-\delta u_{\nu})-2E_{\text{static}}}{\delta u_{\nu}^2}.
\end{equation}

\noindent
Hence two exciton energies at $\pm \delta u_{\nu}$ for
each mode $\nu$, as well as the `static' exciton energy $E_{\text{static}}$ need to be computed. For a molecule with $N$ atoms,
this means that a total of $2\cdot (3N-6)+1$ exciton energy calculations
are required, which for small molecules can be significantly smaller than the number of configurations required to converge a Monte Carlo calculation, a point that we return to in subsection \ref{quadratic_accuracy}.
In principle, $\delta u$ is an infinitesimal quantity that should be as small as possible in the above finite difference formula, however in
practice one needs to choose a finite value for this parameter in order
to avoid numerical divergence issues. 
For perfectly
quadratic energy surfaces the result would be independent of the displacement $\delta u$, and we can therefore conceptually extend the meaning of the 
quadratic formula to simply represent a finite displacement of the system along an individual vibrational mode.
We return to this issue in subsection \ref{quadratic_accuracy}.

Finally, it is worth transforming equation\,\ref{eq:quadratic} into a
different form, which is not used for practical calculations within this
work, but provides valuable intuition for the effects that we discuss. Let us
assume the frequency of vibrational normal mode $\nu$ in the ground state
to be equal to $\omega_{G\nu}^2$. If we further assume that this same normal
mode is present in the excited state with a different frequency $\omega_{E\nu}^2$, then equation\,\ref{eq:quadratic} becomes:

\begin{equation}
    \label{eq:quadratic_equivalent}
E(T)=E_{\text{static}}-\frac{1}{4}\sum_{\nu}\frac{\omega_{G\nu}^2-\omega_{E\nu}^2}{\omega_{G\nu}}[\frac{1}{2}+n_B(\omega_{\nu},T)]+\mathcal{O}(u^4),
\end{equation}

\noindent
This formula is identical to the one
appearing in Ref. \cite{Ferrer2013}, where $E(T)$ is called first moment of the exciton. In
the terms we have used here, the first moment is simply the expectation
value of the exciton energy in the presence of molecular vibrations. 
Equation\,\ref{eq:quadratic_equivalent} demonstrates that
the renormalization of the static exciton energy due to each normal mode
depends on the frequency difference $\omega_{G\nu}^2-\omega_{E\nu}^2$, and
suggests
that $E(T)<E_{\text{static}}$ as long as the potential energy surface of
the ground state has a higher curvature than that of the excited state and therefore 
$\omega_{G\nu}^2>\omega_{E\nu}^2$. This analytical result is essentially the mathematical
manifestation of the intuitive picture of Figure\,\ref{fig:schematic},
and holds in the general multi-dimensional case, at least within the
validity of the quadratic approximation.

\section{Results and discussion}
\label{Results}

\subsection{Comparison to published results}
\label{comparisons}

\begin{figure}[tb]
\centering
\includegraphics[width=0.7\linewidth]{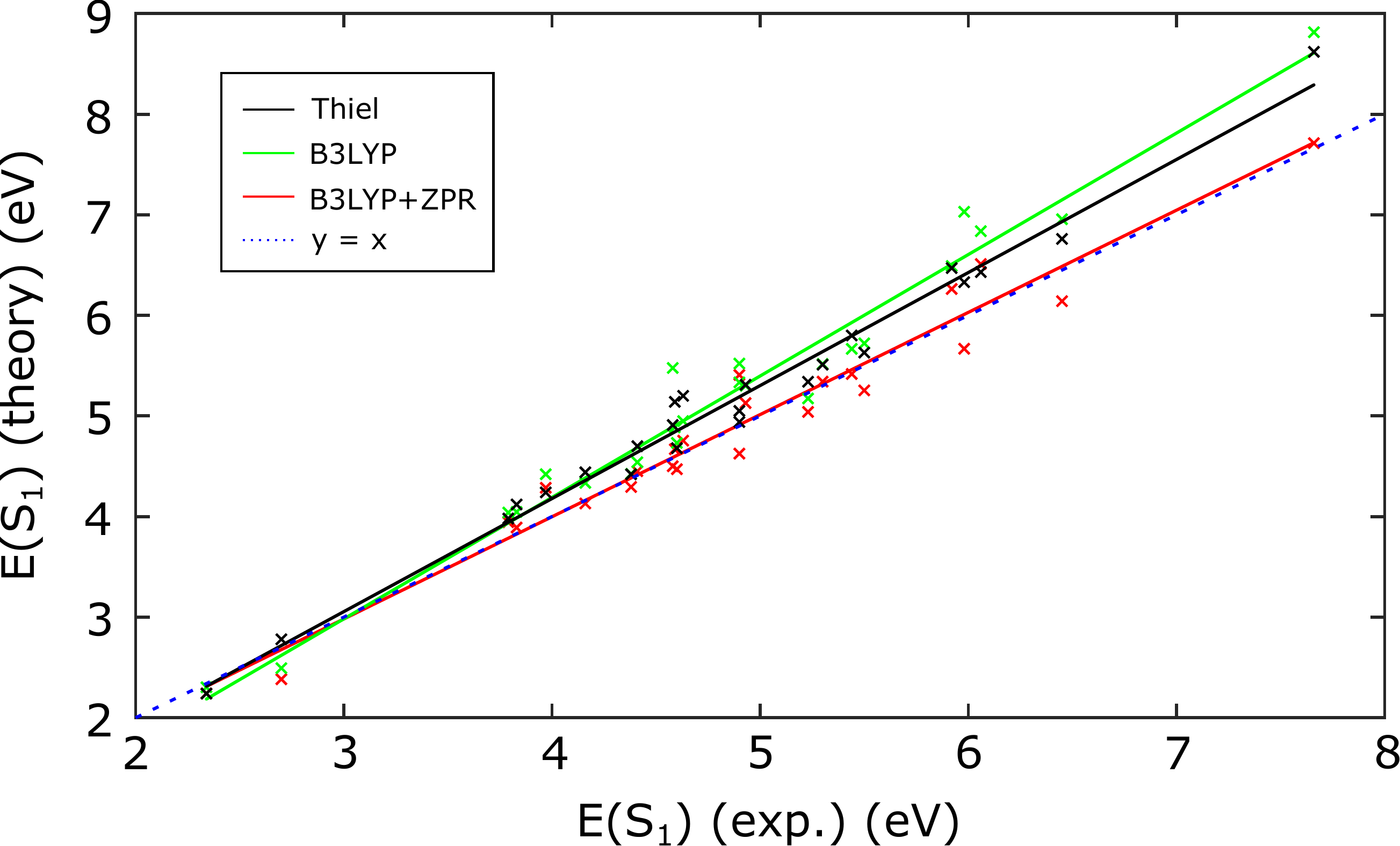}
\caption{Comparison of the different methods used to obtain the singlet exciton energies of the Thiel set to experiment. A `static' TD-DFT approach at the B3LYP level (green) performs the worst, and accurate wavefunction-based methods reported in the literature (black) improve overall agreement (the values in black refer to the CASPT2 Thiel values \cite{Schreiber2008} for single excitations, corresponding to the states computed within TD-DFT here). However, accounting for the zero-point renormalization (ZPR) of exciton energies due to molecular quantum fluctuations leads to the best agreement with experiment (red), even when ignoring beyond-TD-DFT electronic effects.}
\label{fig:comparison}
\end{figure}

We compare the energies of the singlet excitons 
obtained within our Monte Carlo approach by computing exciton energies at
the B3LYP level of TD-DFT and including the effects of zero-point renormalization (ZPR), 
to the experimental values for the corresponding maxima. The references
to the experimental works are summarized in SI section\,S2. These same references were collected in the original publication on the Thiel set \cite{Schreiber2008}, however the values can be slightly different 
for a few of the studied molecules, as we had to ensure that we compare to the band maximum in every case. 
We also compare the values obtained from 
`static' TD-DFT and from highly accurate complete active space second-order
perturbation theory (CASPT2) \cite{Andersson1990} calculations in the 
original Thiel publication \cite{Schreiber2008} to experiment. 
The computed exciton energies always
refer to the first single excitation as described within TD-DFT, and
we exclude any double-excitations (bi-excitons) from our analysis, which
are however accessible using wavefunction-based methods 
\cite{Schreiber2008}. 
In Figure\,\ref{fig:comparison} we plot the computed versus experimental
values for the three approaches, for all the studied molecules for which experimental data is given in Ref. \cite{Schreiber2008}. The closer
a point lies to the $y=x$ line, the better the agreement between theory
and experiment. We fit a linear model to the three sets of results, as a 
means of visualizing the overall agreement with experiment, from where
it becomes evident that the TD-DFT B3LYP+ZPR (red) results provide a significant improvement to the static TD-DFT B3LYP values (green). The numerical results of the Monte Carlo 
simulations for the exciton energy renormalization are summarized in SI 
section\,S1, along with the associated statistical uncertainties. The parameters for all the linear fits are given
in SI section\,S2.
Naturally the Thiel values (black), which were obtained using accurate wavefunction-based methods and capture correlation effects, 
provide an improvement to static TD-DFT. However, from Figure\,\ref{fig:comparison} it is evident that
the correction to TD-DFT exciton
energies induced by the quantum fluctuations of molecular vibrations is at least as significant as their
correction due to the correlation effects included in the Thiel calculations, at least for the specific set of molecules we study 
here. From Figure\,\ref{fig:comparison} it also becomes obvious that the ZPR of exciton energies can become very large, with values that can be as
high as $1.362$\,eV in the case of pyrrole. The average correction is found to be $(345 \pm 67)$\,meV, which is
substantial and certainly needs to be accounted for in order for exciton
energy calculations to achieve predictive accuracy. We have also investigated the effect of increasing the temperature from $0$\,K to $300$\,K, however we found that for the vast majority of molecules the
resulting difference in the exciton energy is small, as summarized in 
SI section S3. This is intuitively obvious from the fact that small organic molecules are dominated by high-frequency vibrational modes such as carbon-carbon stretching motions, which generally lie significantly above the threshold for thermal activation at room temperature. 

\begin{table}[t]
\centering
  \setlength{\tabcolsep}{8pt} 
\begin{tabular}{ccccc}
\hline
\hline
& bias (eV) & rel.bias & RMSE (eV) & rel.RMSE \\
\hline
B3LYP & $0.343$ & $0.0673$ & $0.495$ & $0.091$\\
Thiel & $0.283$ & $0.0545$ & $0.357$ & $0.067$\\
B3LYP+ZPR & $0.011$ & $0.0003$ & $0.231$ & $0.050$\\
\hline
\hline
\end{tabular}
\caption{Statistical measures of the accuracy of the different methods. The average values of the (relative) bias and the (relative) root mean-squared error (RMSE) as defined in the text
are given. The comparisons to experiment refer to the $24$ molecules for which experimental data was found (see SI section\,S2).}
\label{table:statistics}
\end{table}

\begin{figure}[tb]
\centering
\includegraphics[width=0.7\linewidth]{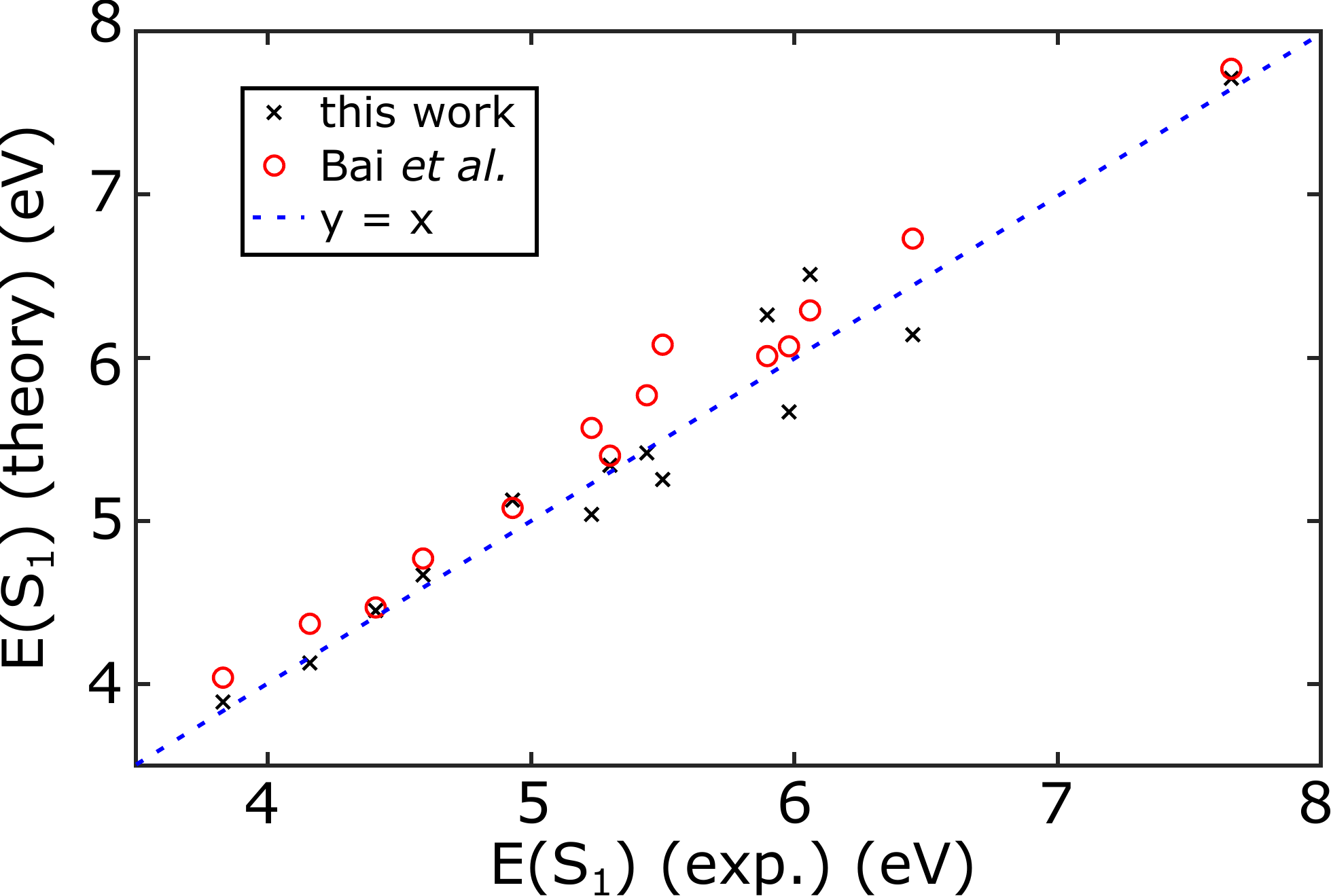}
\caption{Comparison of the exciton energies obtained within our Monte Carlo approach employing TD-DFT and accounting for nuclear quantum effects (B3LYP+ZPR), to the values reported in Ref. \cite{Bai2020} using the nuclear ensemble method in conjunction with coupled cluster calculations. Both methods lead to excellent agreement with experiment.}
\label{fig:comparison_Barbatti}
\end{figure}

\begin{table}[t]
\centering
  \setlength{\tabcolsep}{8pt} 
\begin{tabular}{ccccc}
\hline
\hline
& bias (eV) & rel.bias & RMSE (eV) & rel.RMSE \\
\hline
Bai \emph{et al.} \cite{Bai2020} & $0.211$ & $0.040$ & $0.250$ & $0.047$\\
This work & $0.011$ & $0.003$ & $0.218$ & $0.037$\\
\hline
\hline
\end{tabular}
\caption{Statistical measures of the accuracy of our work within a Monte Carlo (B3LYP+ZPR) approach
and the results reported in \cite{Bai2020} using the nuclear ensemble
method combined with coupled cluster calculations. The average values of 
the (relative) bias and the (relative) root mean-squared error (RMSE) as defined in the text
are given. The comparison is restricted to the structures for which the
same exciton state has been studied in our work and in Ref. \cite{Bai2020}.}
\label{table:statistics_Barbatti}
\end{table}

We employ four different statistical measures in order to rigorously compare the
accuracy of the different methods presented in Figure\,\ref{fig:comparison}, which we summarize in Table\,\ref{table:statistics}. In particular, for our $n=24$
studied molecules with computed exciton energies $b_i$
and experimental exciton energies $a_i$, we compute the
average values of the bias: $\text{bias}=\frac{1}{n}\sum_i (b_i-a_i)$,
the root mean-squared error: $\text{RMSE}=\sqrt{\frac{1}{n}\sum_i (b_i-a_i)^2}$, as well as the relative values to these quantities:
$(\text{rel.bias})=\frac{1}{n}\sum \frac{b_i-a_i}{a_i}$,
$(\text{rel.RMSE}) = \sqrt{\frac{1}{N}\sum_i \frac{(b_i-a_i)^2}{a_i^2}}$.
The reason that we also consider the relative quantities is that
the largest
errors for the static TD-DFT and Thiel approaches are found in the region
of large exciton energies, hence these molecules dominate the 
bias and root mean-squared error. For example, the static 
TD-DFT exciton energy of ethene is renormalized from $8.8$\,eV 
to $7.7$\,eV once we account for nuclear quantum effects.
Using the relative bias and root mean-squared error, which are calculated by dividing by $a_i$ and $a_i^2$ respectively, we achieve a fairer comparison without under-representing the effect of molecules with lower exciton energies. 
We see from Table\,\ref{table:statistics} that the average value
of all the employed statistical measures is 
minimized when using TD-DFT at the B3LYP level combined with the 
zero-point renormalization (B3LYP+ZPR) of exciton energies due to molecular
vibrations. We therefore quantitatively confirm the observation of Figure\,\ref{fig:comparison}: for the studied set of molecules, corrections to
static exciton energies due to nuclear quantum motion are generally larger
than corrections due to beyond-TD-DFT electronic effects that are included in the Thiel values. We also quantify the spread of
the four statistical quantities around their average value by
computing their standard deviation, which is given in SI Table\,S32.
We find that the spread of the B3LYP+ZPR values is in every case
comparable to that of the Thiel values, and significantly smaller
than that of the static 
TD-DFT (B3LYP) values. 

Similar results to ours were presented in Ref. \cite{Bai2020} using the 
nuclear ensemble method and coupled cluster calculations, although the 
computed values were not compared to experiment in that case. In Figure\,\ref{fig:comparison_Barbatti} we compare our
results obtained within the Monte Carlo method at the TD-DFT level to
those of Ref. \cite{Bai2020}. We restrict the comparison to fourteen structures for which the
same exciton state has been studied, as Ref. \cite{Bai2020} also reports
the energies of some excitons that are not necessarily the lowest-lying 
ones, which we chose to study in this work.
We see that both computational
methods lead to excellent agreement with experiment, and a rigorous
comparison is made in Table\,\ref{table:statistics_Barbatti}. Among the
reported statistical measures, the RMSE and relative RMSE are the ones
that most accurately capture the accuracy of the two methods, as they do
not suffer from effects such as cancellations of errors, which are 
evidently present in the bias and relative bias when using TD-DFT due to 
some values underestimating and some others overestimating experiment. This
is not a problem with coupled cluster calculations in the studied 
molecules, since these generally predict an exciton energy value that
is higher compared to TD-DFT, which also becomes apparent from the fact
that the values in Ref. \cite{Bai2020} tend to slightly overestimate
experiment, as seen in Figure\,\ref{fig:comparison_Barbatti}. 
Since the agreement of our results within TD-DFT to experiment is comparable to the agreement of the results of Bai \emph{et al.} which were obtained using coupled cluster calculations, it is once again highlighted that 
at least for the specific set of studied molecules, the ZPR of exciton energies provides a larger, or at least comparable, correction to the static TD-DFT exciton energies compared to the correction obtained by using
more accurate electronic structure methods. Of course it is still generally
the case that one needs both an accurate description of the electronic
structure and of vibrational effects in order to achieve better quality
results. To demonstrate this fact,
we have chosen four molecules for which TD-DFT calculations at the B3LYP
level provide a poor starting point for the static exciton energy, as well 
as inaccurate values once ZPR is accounted for, and we now perform a Monte
Carlo sampling by employing significantly more accurate coupled cluster singles and doubles (CCSD) calculations. These molecules are benzoquinone, cyclopropene, formamide, and pyrrole, and the results are summarized
in Table\,\ref{table:improvement_ccsd}. In all cases, the corrected exciton
energy obtained by using CCSD calculations within our Monte Carlo sampling
is closer to the experimental value compared to the TD-DFT case. However, even in this case the experimental values for benzoquinone
and cyclopropene are overestimated by $0.2-0.3$\,eV. We attribute this behavior to the still limited treatment
of correlations within CCSD compared to more accurate methods, which results in these calculations commonly
overestimating excited state energies as also observed in the results
of Ref.\,\cite{Bai2020} shown in Figure\,\ref{fig:comparison_Barbatti}.

\begin{table}[t]
\centering
  \setlength{\tabcolsep}{8pt} 
\begin{tabular}{cccccc}
\hline
\hline
molecule & B3LYP & B3LYP+ZPR & CCSD & CCSD+ZPR & experiment \\
\hline
benzoquinone & $2.493$ & $2.383(15)$ & $3.126$ & $2.999(15)$ & $2.7$ \cite{Weber2001} \\
cyclopropene & $6.956$ & $6.141(30)$ & $7.508$ & $6.663(32)$ & $6.45$ \cite{Robin1969}\\
formamide & $5.722$ & $5.254(57)$ & $5.876$ & $5.456(56)$ & $5.5$ \cite{Clark1995}\\
pyrrole & $7.030$ & $5.668(43)$ & $6.822$ & $5.935(78)$ & $5.98$ \cite{Flicker1976}\\
\hline
\hline
\end{tabular}
\caption{Comparison of the energy of the first excited state computed at
the different levels of theory to experimental values. For the cases where zero-point renormalization (ZPR) is computed, the associated statistical uncertainty of the final exciton energy is given in parenthesis. }
\label{table:improvement_ccsd}
\end{table}

\subsection{Exciton energy renormalization at different levels of theory}
\label{levels_of_theory}

Ideally, one would combine accurate electronic structure calculations with our approach for 
accounting for nuclear quantum motion. However, for most of the studied
molecules we found that a minimum of approximately $50$ points is
required to converge the Monte Carlo sampling of the integral in equation\,\ref{eq:harm_exp_value}. While the cost of $50$ TD-DFT calculations of small molecules is reasonably low, it quickly becomes
very high as one moves to more accurate wavefunction-based methods. 
It is therefore reasonable to wonder whether the vibrational 
corrections to exciton energies could be computed at one 
(cheaper) level of theory and applied to the static exciton
energy obtained at another (more accurate) level of theory.
Some first conclusions on this can be drawn from the comparison of the excited state energy corrections at the TD-DFT B3LYP and CCSD
levels of Table\,\ref{table:improvement_ccsd}; with the exception of pyrrole for which the B3LYP correction is $1.5$ times larger than the
CCSD one, the maximum deviation of the B3LYP zero-point renormalization from the CCSD value is $13\%$. 
While of course this is encouraging, these data points alone are not enough to draw a general conclusion. 
However, given this good agreement between the few B3LYP and CCSD excited state energy
corrections that we compared, the fact that hybrid functionals
such as B3LYP are known to lead to an accurate description
of the electron-vibration interactions in organic molecules \cite{LaflammeJanssen2010},  and also the excellent agreement between theory and experiment in Figure\,\ref{fig:comparison}, it seems
a reasonable conclusion that hybrid functionals provide an accurate description of exciton-vibration coupling and the zero-point
renormalization this induces.

In order to check whether the vibrational correction to exciton energies
changes when using different electronic structure methods in a more systematic manner, we repeated our Monte Carlo sampling
using the local density approximation (LDA) and pure Hartree-Fock (HF) exchange
within TD-DFT, while using the
vibrational modes obtained at the
B3LYP level. This choice of functionals allows us to comment on the role
of exact exchange for exciton-vibration interactions: in particular, LDA
includes no exact exchange, while B3LYP includes $20\%$ and HF represents
the case of including full exact exchange. Since the included fraction of exchange is generally
known to have little effect on the structure of a molecule and the computed
ground-state vibrational modes \cite{LaflammeJanssen2010}, using the B3LYP geometries and vibrational modes
is a reasonable approximation. Therefore, all differences in the values of
the exciton energy renormalization between the different functionals are
purely due to the variations of the excited state wavefunction as obtained
from TD-DFT.

The numerical results of the calculations employing the different functionals are presented
in SI section\,S1. 
We compare the LDA and HF energy
corrections to the B3LYP ones and find that on average, LDA predicts
a red-shift of the excited state energy which is $1.3$ times greater than
B3LYP, and HF $1.8$ times greater. This is in agreement
with reports on electron-vibration interactions being strongly affected by the fraction of exact exchange included in the calculation \cite{LaflammeJanssen2010}. The results suggest that LDA provides a reasonable approximation to B3LYP for computing zero-point renormalization
in the studied organic molecules, with only a few structures showing very strong deviations. As shown in SI Figure\,S1, the largest
differences between the corrections predicted by LDA and B3LYP are found for smaller molecules with less than $60$ electrons. We attribute
this observation to the fact that the excited states of these smaller structures have a greater electronic density in the vicinity of
the localized high-frequency vibrations that can dominate in organics, hence leading to a very strong coupling that is very sensitive to
the detailed structure of the excited state wavefunction. These points are revisited in subsections\,\ref{molecular_size} and \ref{mode_resolved} where the impact of molecular size and individual vibrational modes on zero-point renormalization are discussed
respectively.

Overall, we find that in most organic structures, hybrid functionals such as B3LYP provide a good approximation to the values of
zero-point renormalization computed with more accurate methods such as CCSD. For several of the studied molecules, even the complete omission of exact exchange through the use of LDA seems to also give reasonable results within $30\%$ of the B3LYP values. Most strong deviations between electronic structure
methods are found for smaller molecules, which is in a sense encouraging, since these are the ones that can commonly be studied using more 
expensive methods without too great of a computational cost.

\subsection{The impact of the molecular size}
\label{molecular_size}

\begin{figure}[tb]
\centering
\includegraphics[width=\linewidth]{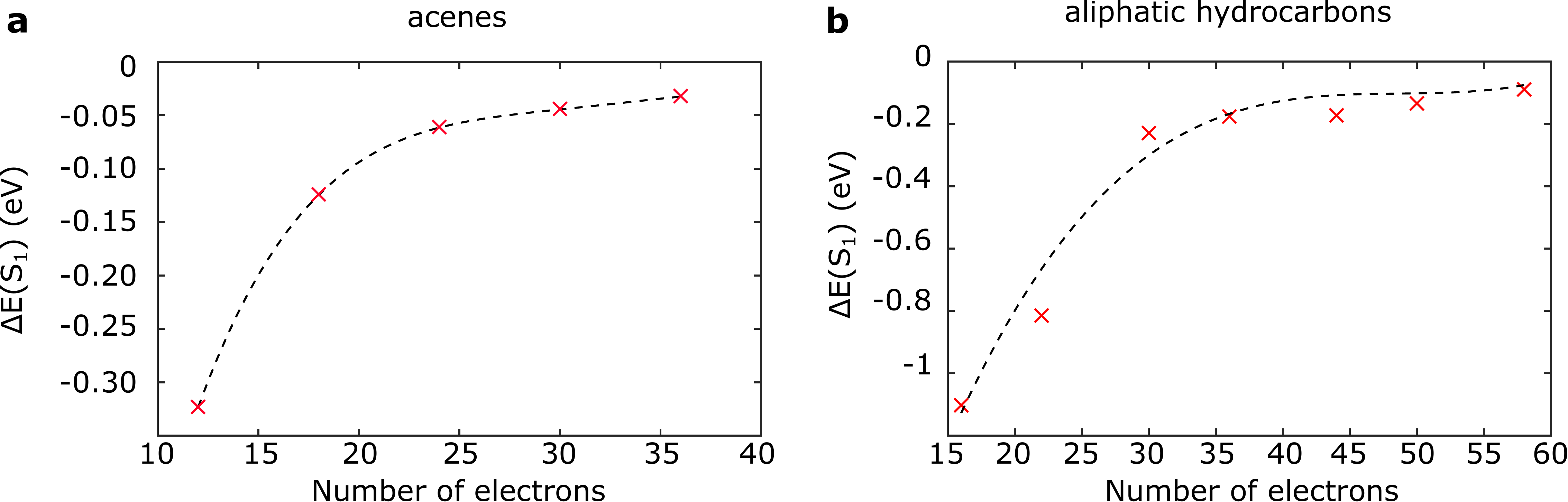}
\caption{Size dependence of the correction $\Delta \text{E}(\text{S}_1)$ that nuclear quantum motion induces to the singlet exciton energies  of the acene (panel \textbf{a}) and aliphatic hydrocarbon (panel \textbf{b}) families of molecules.}
\label{fig:size}
\end{figure}

We found in Ref.\,\cite{Alvertis2020} for periodic molecular crystals that the correction
to exciton energies due to nuclear quantum motion becomes more important
for smaller molecules. For the single-molecule systems studied here
we find that the same trend holds, as shown in Figure\,\ref{fig:size}
for the acene and aliphatic hydrocarbons families of molecules. The results 
in Figure\,\ref{fig:size} are obtained at the TD-DFT B3LYP level of 
theory.
We believe that the reason for the observed trend is that
smaller molecules have a greater electronic density in the vicinity of
localized atomic motions that are activated due to quantum fluctuations,
hence leading to stronger exciton-vibration interactions. Another
way of reaching the same conclusion is using H\"{u}ckel theory.
Let us consider the example of linear polyenes, such as
ethene, butadiene, hexatriene and octatetraene that are studied
here and are included among other aliphatic hydrocarbons in 
Figure\,\ref{fig:size}b, showing a decrease in the exciton energy
correction with increasing system size. For a linear polyene
consisting of $N$ carbon
atoms, H\"{u}ckel theory predicts
the energy of the $i^{\text{th}}$ orbital
is: $E_i=\alpha+2\beta \cos{(2\pi n/(N+1))}$, with $\alpha,\beta$ the
H\"{u}ckel parameters, and $i=1,2,...,N$. From this, we find
for the HOMO-LUMO gap:

\begin{equation}
    \label{eq:Huckel_gap}
    E_{\text{gap}}\propto \sin(\frac{\pi}{2(N+1)}).
\end{equation}

\noindent
In the limit of $N \rightarrow \infty$, this gives 
$E_{\text{gap}} \propto 1/N$. 
A similar result can be obtained
for cyclic polyenes. Assuming that the same holds for any general
conjugated molecule, this suggests that in larger systems 
the HOMO is less bonding and the LUMO less
anti-bonding than in comparably smaller systems. Therefore,
from the intuitive picture of Figure\,\ref{fig:schematic}, the curvature of the ground and excited state surfaces
of large molecules is similar, and the difference in their normal
mode frequencies is small. Hence, according to the
expression\,\ref{eq:quadratic_equivalent} that results from the quadratic approximation, the correction to the exciton energy
is also smaller than that for a smaller molecule. This is encouraging, in the sense that corrections
due to nuclear quantum motion are mostly important for small systems,
for which they are cheaper to compute. 

\subsection{Accuracy and speed of the quadratic method}
\label{quadratic_accuracy}

\begin{figure}[tb]
\centering
\includegraphics[width=0.7\linewidth]{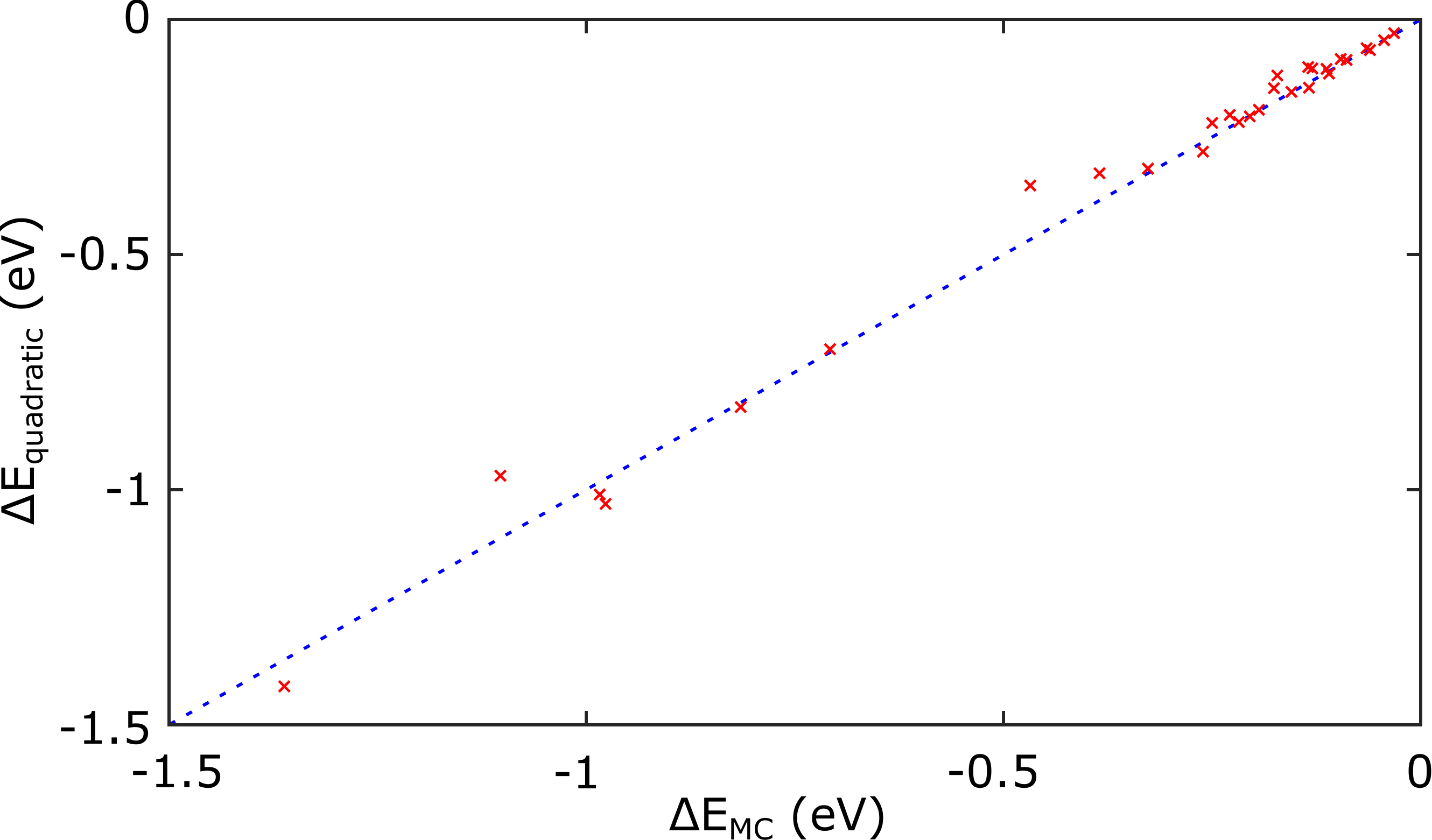}
\caption{Comparison of values predicted by the quadratic and Monte Carlo
methods for the zero-point renormalization of the exciton
energy of the different molecules. The $y=x$ line is given in blue for reference.}
\label{fig:comparison_MC_quadratic}
\end{figure}

Having established the accuracy of the Monte Carlo method for computing
the exciton energy renormalization due to nuclear quantum motion, we now proceed to
use it as a benchmark to assess the accuracy of the quadratic method. Figure\,\ref{fig:comparison_MC_quadratic} visualizes the quadratic versus
the Monte Carlo correction for each of the studied molecules, using TD-DFT and the B3LYP functional, along with the cc-pVDZ basis set. The better
the agreement between the two methods for a given molecule, the closer
the associated point lies to the $y=x$ line (dashed line). Overall, the
agreement between the two methods is good, and we propose that the quadratic method can indeed be used to make quantitative predictions of
exciton energies including vibrational renormalization effects. Particularly for smaller molecules, this has the additional advantage that it comes at a lower computational
cost compared to Monte Carlo sampling (see also theoretical background section and the discussion accompanying Figure\,\ref{fig:comparison_CPU_hours} below).
The converged values for the exciton energies as predicted by the quadratic
method are summarized in SI section\,S5. 

\begin{figure}[tb]
\centering
\includegraphics[width=\linewidth]{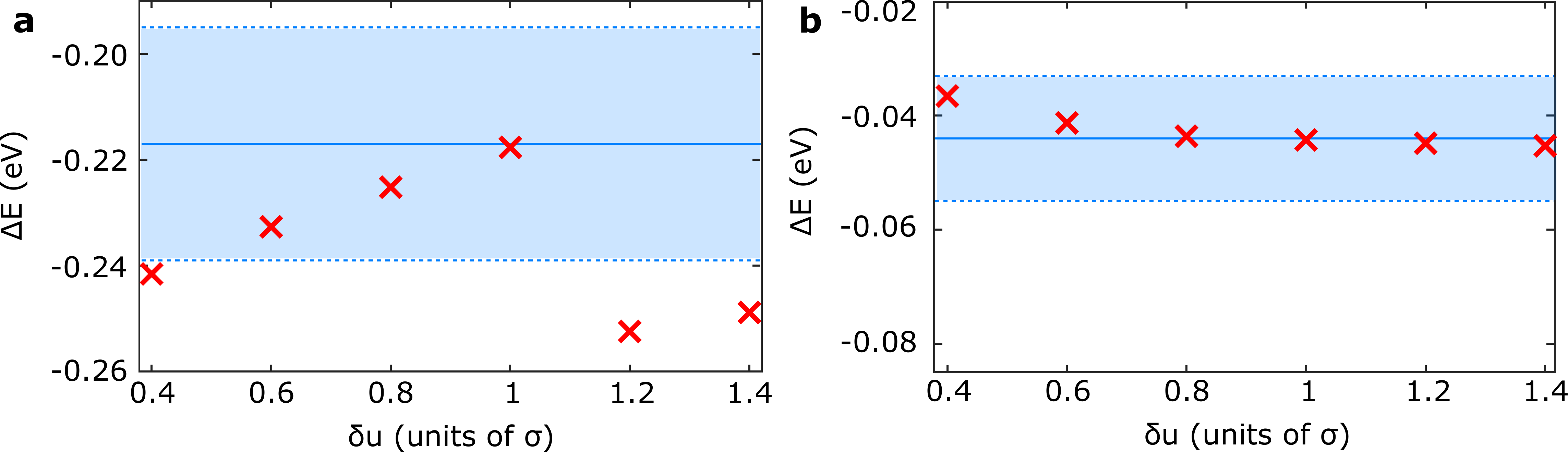}
\caption{Zero-point renormalization $\Delta E$ of the exciton energy predicted by the quadratic method, with respect to the displacement $\delta u$ appearing in equation\,\ref{eq:quadratic}), for the case of pyrazine (panel \textbf{a}) and tetracene (panel \textbf{b}). The energy shift predicted by the Monte Carlo approach is given for reference (blue line) and its statistical uncertainty indicated with the blue shaded region.}
\label{fig:quadratic_convergence}
\end{figure}

As already mentioned in subsection \ref{quadratic}, when computing
the quadratic correction to the exciton energies, one needs to make a 
choice for the displacement $\delta u$ appearing in 
equation\,\ref{eq:quadratic}. We find that generally a displacement of
$\delta u = \sigma$, where $\sigma$ is the standard deviation of the thermal quantum distribution appearing in
equation\,\ref{eq:Gaussian_width}, leads to results that are
in very close agreement with the computed Monte Carlo values. 
Figures\,\ref{fig:quadratic_convergence}a and 
\ref{fig:quadratic_convergence}b show the quadratic 
correction of pyrazine and tetracene respectively, comparing the quadratic
values at different values of $\delta u$ (red crosses) with the Monte Carlo
correction (blue line) and its associated statistical uncertainty (blue shaded region). The
variation of the quadratic correction due to changes in $\delta u$ is
generally comparable to the statistical uncertainty of the Monte Carlo method. 

We also explore the computational cost of the quadratic method and compare
to that of a Monte Carlo sampling. In Figure\,\ref{fig:comparison_CPU_hours} we visualize the CPU hours required
to compute the correction to the static exciton energies that is induced
by the zero-point motion of the different molecules. For the Monte Carlo sampling, the calculation always includes taking an average over $100$
configurations as mentioned previously, which we find leads to converged
results and small statistical uncertainties to the final exciton energies (SI section\,S4). Overall,
the quadratic method is cheaper to use for smaller structures with fewer
electrons, where we also argue that the correction to exciton energies
due to vibrations is larger (see Figure\,\ref{fig:size} and discussion). Therefore, given also the accuracy of this method and the
mode-resolved information it provides (see subsection \ref{mode_resolved}), we propose
that it is the better alternative once one approaches the limit of smaller molecules.
For larger structures with several vibrational normal modes we find that
the Monte Carlo method provides a faster estimate of the red-shift and should thus be preferred. 

\begin{figure}[tb]
\centering
\includegraphics[width=0.7\linewidth]{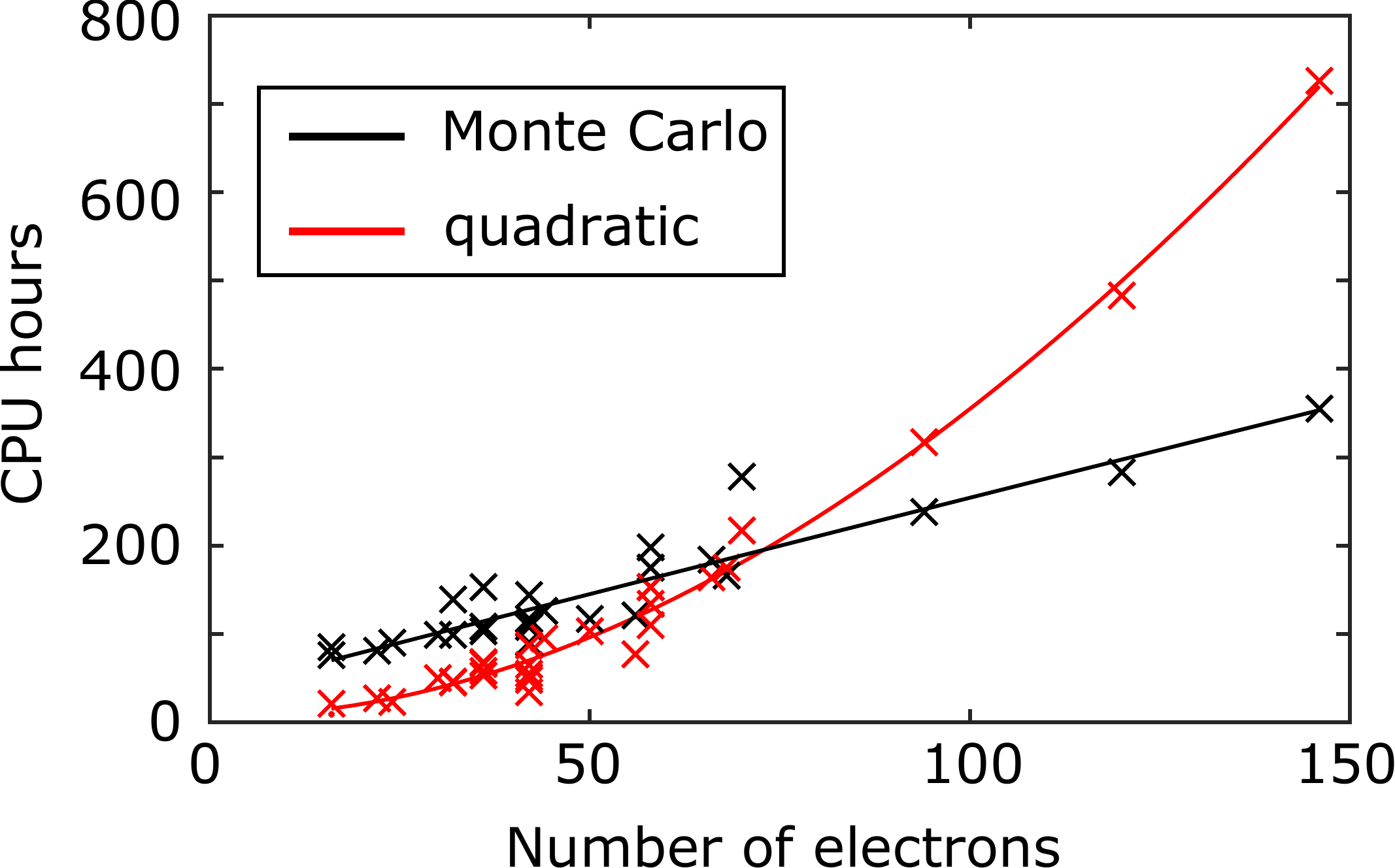}
\caption{Comparison of the CPU hours required to compute the exciton energy correction of the different molecules due to nuclear quantum motion, using the Monte Carlo (black crosses) and quadratic (red crosses) methods. The solid lines are given as guides to the eye.}
\label{fig:comparison_CPU_hours}
\end{figure}

\subsection{Mode-resolved picture for the exciton energy renormalization}
\label{mode_resolved}

\begin{figure}[tb]
\centering
\includegraphics[width=\linewidth]{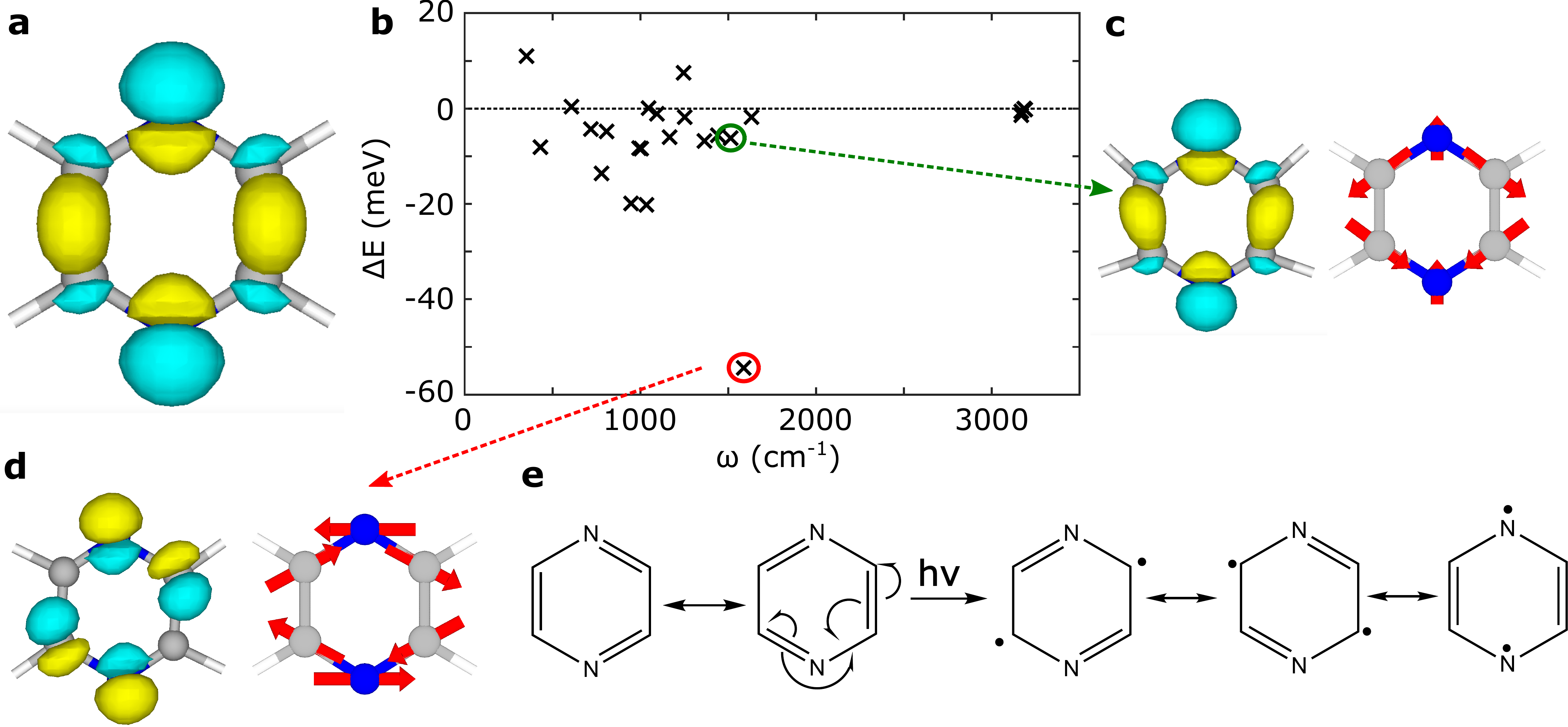}
\caption{Mode-resolved zero-point renormalization of the exciton energy of pyrazine. The renormalization of the exciton energy due to the various normal modes of vibration is given in panel \textbf{b}, with a weakly  coupled mode highlighted in green (displacement pattern in panel \textbf{c}), and the most strongly coupled mode highlighted in red  (displacement pattern in panel \textbf{d}). The exciton wavefunction (transition density) is visualized in panel \textbf{a} in the optimized geometry, and along the weakly and strongly coupled modes in panels \textbf{c} and \textbf{d} respectively. The resonance structures of pyrazine in the ground state and upon photoexcitation are shown in panel \textbf{e}.}
\label{fig:pyrazine_mode_resolved}
\end{figure}

\begin{figure}[tb]
\centering
\includegraphics[width=\linewidth]{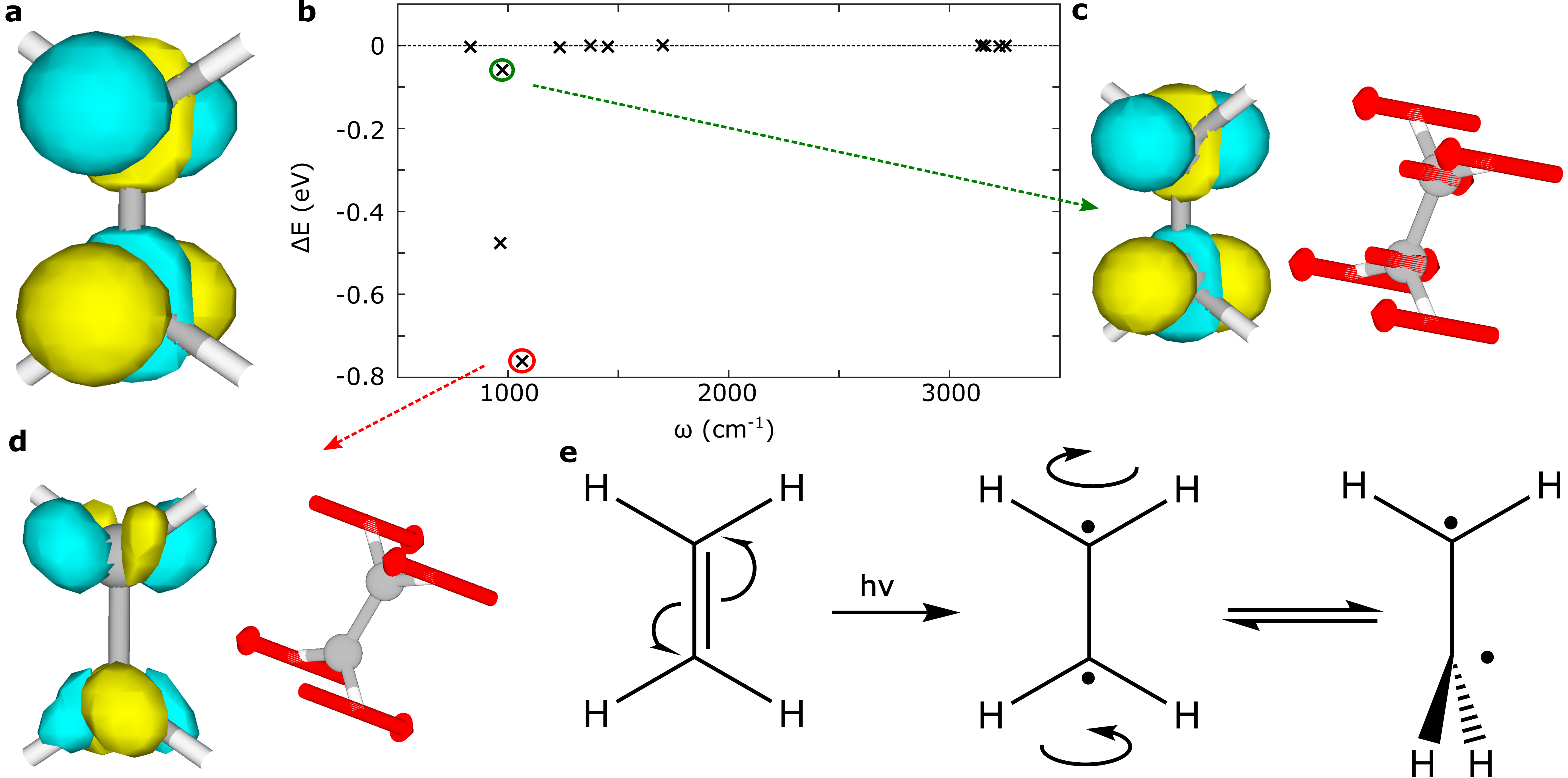}
\caption{Mode-resolved zero-point renormalization of the exciton energy of ethene. The renormalization of the exciton energy due to the various normal modes of vibration is given in panel \textbf{b}, with a weakly  coupled mode highlighted in green (displacement pattern in panel \textbf{c}), and the most strongly coupled mode highlighted in red  (displacement pattern in panel \textbf{d}). The exciton wavefunction (transition density) is visualized in panel \textbf{a} in the optimized geometry, and along the weakly and strongly coupled modes in panels \textbf{c} and \textbf{d} respectively. The resonance structures of ethene in the ground state and upon photoexcitation are shown in panel \textbf{e}.}
\label{fig:ethene_mode_resolved}
\end{figure}

While the quadratic method for computing the renormalization of exciton
energies due to quantum fluctuations is in principle less accurate than performing
a Monte Carlo sampling, it is a cheaper and simpler method that is easier to
implement computationally. However, perhaps its most important advantage
is the fact that it separates the contributions of the individual 
vibrational normal modes to the exciton energy renormalization, as revealed
by equation\,\ref{eq:quadratic}. 
In 
Figures\,\ref{fig:pyrazine_mode_resolved}b and 
\ref{fig:ethene_mode_resolved}b we visualize the mode-resolved 
renormalization of the exciton energy of pyrazine and ethene respectively, at $T=0$\,K.
From these figures, it becomes evident that for both molecules, it is only
a few modes that dominate the zero-point renormalization of the exciton
energy. In both cases, the most important mode for the exciton energy renormalization is highlighted in a red circle. For pyrazine, this dominant mode is responsible for $35$\% of the exciton energy renormalization, while for ethene for $56$\%. 

In order to gain a deeper microscopic understanding of the effect of the
specific vibrational modes on the exciton states of these molecules,
we plot the undistorted `static' exciton wavefunctions (represented by the
transition density) in Figures\,\ref{fig:pyrazine_mode_resolved}a and 
\ref{fig:ethene_mode_resolved}a respectively, together with the 
wavefunction at a typical distorted configuration ($\delta u=\sigma$) along
these dominant modes, visualized in Figures 
\,\ref{fig:pyrazine_mode_resolved}d and \ref{fig:ethene_mode_resolved}d. 
For comparison, we also plot the exciton wavefunction when the molecule
is displaced along a mode which is weakly coupled to the exciton in 
Figures\,\ref{fig:pyrazine_mode_resolved}c and 
\ref{fig:ethene_mode_resolved}c, together with the displacement pattern
of these motions. These weakly coupled vibrational modes are highlighted 
in green in Figures\,\ref{fig:pyrazine_mode_resolved}b and 
\ref{fig:ethene_mode_resolved}b. It is evident that while the displacement of the dominant modes leads to a significant change of the 
excitonic wavefunction from its form at the equilibrium geometry, this is
not the case for weakly coupled modes.

The dominating modes for exciton renormalization can sometimes be qualitatively explained from a chemically intuitive perspective by considering the conventional skeletal structures of the ground and excited state molecules, and (from equation\,\ref{eq:quadratic_equivalent}) which stretching modes are likely to be far weaker (lower frequency) in the excited state compared to the ground state.
For pyrazine in Figure\,\ref{fig:pyrazine_mode_resolved}, as an aromatic molecule similar to benzene, it has two principal resonance structures as shown in Figure\,\ref{fig:pyrazine_mode_resolved}e. The HOMO and LUMO are both $\pi$ orbitals and excitation to $\text{S}_1$ can be qualitatively drawn as breaking a $\pi$ bond, leading to a (singlet) biradical. The biradical \cite{Stuyver2019} has numerous possible resonance structures, three of which are drawn in Figure\,\ref{fig:pyrazine_mode_resolved}e. These quinoidal structures suggest that in the $\text{S}_1$ state, for a given nitrogen it will be easier to elongate one C-N bond while shortening the other to form a quinoidal-like biradical. This corresponds to the dominant stretching vibration in Figure\,\ref{fig:pyrazine_mode_resolved}d.

Similarly, for ethene in Figure\,\ref{fig:ethene_mode_resolved},  in the ground $\text{S}_0$ state the molecule is planar due to the $\pi$ bond, although this causes some small steric hindrance between hydrogen atoms of different carbons. Upon exciting to the $\text{S}_1$ state, the $\pi$ bond is broken, such that it is much easier to rotate around the central C-C bond as shown in Figure\,\ref{fig:ethene_mode_resolved}e and we would expect the frequency of this mode to drastically reduce, leading to a large renormalization by equation\,\ref{eq:quadratic_equivalent}. This rotation is the dominating mode as shown in Figure\,\ref{fig:ethene_mode_resolved}d.

It is
not always true that a subset of modes can be identified to dominate 
the exciton energy renormalization, however some patterns can still be observed. Perhaps the most
obvious one is a ring-breathing mode of frequency in the vicinity of $1600$\,$\text{cm}^{-1}$, which is largely responsible for the exciton energy renormalization in several cyclic compounds, which can be intuitively understood as in the case of pyrazine in Figure\,\ref{fig:pyrazine_mode_resolved}e, where this motion contributes
$35\%$ of the renormalization. Similarly for other cyclic structures:
benzoquinone ($20$\%), pyridine ($38$\%), pyrimidine ($39$\%), tetrazine ($62$\%) and triazine ($30$\%). 
For each of the studied molecules, we give the percentage of contribution
of the mode that most strongly couples to the exciton energy renormalization, along with the frequency of this motion, in SI section\,S5. Overall, in most molecules high-frequency modes dominate
the renormalization of exciton energies, with the average frequency
of the most strongly coupled mode being $\bar{\omega}_{\text{dominant}}=(1024 \pm 121$)\,$\text{cm}^{-1}$. This dominance of high-frequency
modes
is consistent with our
observation of subsection\,\ref{molecular_size} that zero-point renormalization is, for the molecules studied here, more prominent in smaller molecular structures. The
excited states of these systems have a greater electronic density in the vicinity of localized high-frequency motions, such as carbon-carbon stretches, hence the corresponding wavefunctions show stronger variations upon displacement of the atoms. It is for the same reasons that larger differences in the computed energy corrections between the different functionals tend to appear for smaller molecular
structures, as was seen in subsection\,\ref{levels_of_theory}; the strong exciton-vibration interactions in these systems
can be very sensitive to the local structure of the excited state wavefunctions, which depends
on the level of electronic structure theory.

\section{Conclusions}
\label{conclusions}

In this work we have presented an intuitive picture for the red-shift
of exciton energies that is caused by nuclear quantum motion, and two
computational approaches for estimating its magnitude. We find that this red-shift can
indeed be very substantial, reaching values of more than $1$\,eV in some cases.
We rigorously compare our results for the exciton energies of a large set of molecules, obtained within a Monte Carlo framework and time-dependent
density functional theory, to experiment and previous computational
studies, and find that accounting for the effects of nuclear quantum motion in this manner leads to predictive accuracy for exciton energies. 
We show that the magnitude of the red-shift provided by quantum fluctuations is typically larger for smaller molecules, a result which we
explain within a H\"{u}ckel theory picture. Moreover, we find that while
the predicted red-shift caused by vibrations depends on the level of theory
used for the exciton calculations, TD-DFT employing hybrid functionals should provide a good compromise
between accuracy and computational cost for estimating its magnitude.
Naturally, further improvement of the results can be
achieved by combining our Monte Carlo approach with more accurate methods
of electronic structure calculations, such as coupled cluster, as we also demonstrate for a subset of the studied molecules.
Additionally, we employ a quadratic approximation for computing the magnitude
of exciton energy renormalization due to zero-point motion, which allows us to
disentangle the contribution of individual normal modes to this effect. 
While this method is conceptually easier to implement compared
to a Monte Carlo sampling, it is 
in principle less accurate as well. Nevertheless, we find that it performs
very well for the diverse set of molecules which are studied here, and it 
also provides a cheaper way of computing the renormalization of exciton 
energies in smaller molecules, where this effect is also most relevant. By
using the quadratic method we find that for several molecules the 
renormalization of the exciton energy is dominated by a few normal modes
of vibration, with a ring-breathing motion playing a prominent role in the red-shift of the exciton energy in several cyclic compounds.
Overall, our study provides critical microscopic insights into the effect
of nuclear quantum motion on exciton energies at equilibrium, and emphasizes
its importance for achieving predictive accuracy.

\section*{Supplementary Material}
See supplementary material for the full results of the Monte Carlo sampling using
different functionals, details for the comparison of these results to experiment, convergence tests and results at $300$\,K. Detailed data on the
vibrational modes dominating the quadratic correction of the exciton energy
are also provided.

\section*{Data availability}
The data that support the findings of this study are openly available in
Apollo, the University of Cambridge repository, at \url{https://doi.org/10.17863/CAM.70585}, reference number \cite{data}.

\section*{Acknowledgments}
The authors acknowledge useful discussions with Stuart Althorpe (Cambridge) and the anonymous reviewers for their constructive feedback.  T.J.H.H. acknowledges a Royal Society University Research Fellowship Ref: URF{\textbackslash}R1\hspace{-0.3mm}{\textbackslash}201502. B.M. acknowledges support from the Gianna Angelopoulos Programme for Science, Technology, and Innovation. A.M.A. acknowledges the support of the Engineering and
Physical Sciences Research Council for funding under grant 
EP/L015552/1 and the Winton Programme for the Physics of Sustainability. Part of the 
calculations were performed using resources provided by the Cambridge Tier-2
system operated by the University of Cambridge Research Computing Service 
(http://www.hpc.cam.ac.uk) and funded by EPSRC Tier-2 capital grant
EP/P020259/1.

\textbf{}
\bibliographystyle{unsrt}
\bibliography{references}

\end{document}


\maketitle
\tableofcontents

\clearpage

\section{Monte Carlo exciton energies at the different levels of theory}

We compute the static and zero-point corrected exciton energies
for the different studied molecules within three different levels
of TD-DFT: 1. the local density approximation (LDA), 2. using the hybrid B3LYP functional \cite{Becke1993} and 3. using pure Hartree-Fock (HF)
exchange. Tables\,\ref{table:acetamide}-\ref{table:pentacene}
give the static and corrected exciton energies at these
different levels of theory for the studied molecules, as well
as the corrections induced by nuclear quantum motion. The statistical uncertainty (standard
error) to the zero-point renormalization, as obtained from the Monte Carlo sampling using $100$ configurations at $T=0$\,K, is given (in meV) in parentheses next to its value.

\begin{table}[t]
\centering
  \setlength{\tabcolsep}{12pt} 
  \renewcommand{\arraystretch}{1.5} 
\begin{tabular}{cccc}
\hline
\hline
DFT level & LDA & B3LYP & HF \\
\hline
$E(\text{S}_1)_{\text{static}}$ (eV) & $5.479$ & $5.667$ & $6.671$\\
$E(\text{S}_1)_{\text{ZPR}}$ (eV) & $5.1$ & $5.417$ & $6.51$\\
$\Delta E(\text{S}_1)$ (eV) & $-0.379(39)$ & $-0.250(36)$ & $-0.161(29)$\\
\hline
\hline
\end{tabular}
\caption{Energy of the first singlet exciton of acetamide at the different levels of theory and with/without corrections due to zero-point renormalization.}
\label{table:acetamide}
\end{table}

\begin{table}[t]
\centering
  \setlength{\tabcolsep}{12pt} 
  \renewcommand{\arraystretch}{1.5} 
\begin{tabular}{cccc}
\hline
\hline
DFT level & LDA & B3LYP & HF \\
\hline
$E(\text{S}_1)_{\text{static}}$ (eV) & $4.259$ & $4.424$ & $5.152$\\
$E(\text{S}_1)_{\text{ZPR}}$ (eV) & $4.097$ & $4.294$ & $5.085$\\
$\Delta E(\text{S}_1)$ (eV) & $-0.162(27)$& $-0.130(27)$ & $-0.067(23)$\\
\hline
\hline
\end{tabular}
\caption{Energy of the first singlet exciton of acetone at the different levels of theory and with/without corrections due to zero-point renormalization.}
\label{table:acetone}
\end{table}

\begin{table}[t]
\centering
  \setlength{\tabcolsep}{12pt} 
  \renewcommand{\arraystretch}{1.5} 
\begin{tabular}{cccc}
\hline
\hline
DFT level & LDA & B3LYP & HF \\
\hline
$E(\text{S}_1)_{\text{static}}$ (eV) & $4.181$ & $4.949$ & $6.349$\\
$E(\text{S}_1)_{\text{ZPR}}$ (eV) & $4.094$ & $4.755$ & $6.102$\\
$\Delta E(\text{S}_1)$ (eV) & $-0.087(31)$ & $-0.194(27)$ & $-0.247(25)$\\
\hline
\hline
\end{tabular}
\caption{Energy of the first singlet exciton of adenine at the different levels of theory and with/without corrections due to zero-point renormalization.}
\label{table:adenine}
\end{table}

\begin{table}[t]
\centering
  \setlength{\tabcolsep}{12pt} 
  \renewcommand{\arraystretch}{1.5} 
\begin{tabular}{cccc}
\hline
\hline
DFT level & LDA & B3LYP & HF \\
\hline
$E(\text{S}_1)_{\text{static}}$ (eV) & $1.838$ & $2.493$ & $5.31$\\
$E(\text{S}_1)_{\text{ZPR}}$ (eV) & $1.752$ & $2.383$ & $3.885$\\
$\Delta E(\text{S}_1)$ (eV) & $-0.086(13)$ & $-0.110(15)$ & $-1.425(15)$\\
\hline
\hline
\end{tabular}
\caption{Energy of the first singlet exciton of benzoquinone at the different levels of theory and with/without corrections due to zero-point renormalization.}
\label{table:benzoquinone}
\end{table}

\begin{table}[t]
\centering
  \setlength{\tabcolsep}{12pt} 
  \renewcommand{\arraystretch}{1.5} 
\begin{tabular}{cccc}
\hline
\hline
DFT level & LDA & B3LYP & HF \\
\hline
$E(\text{S}_1)_{\text{static}}$ (eV) & $6.445$ & $6.491$ & $6.635$\\
$E(\text{S}_1)_{\text{ZPR}}$ (eV) & $5.965$ & $6.262$ & $6.46$\\
$\Delta E(\text{S}_1)$ (eV) & $-0.480(21)$ & $-0.229(21)$ & $-0.175(32)$\\
\hline
\hline
\end{tabular}
\caption{Energy of the first singlet exciton of butadiene at the different levels of theory and with/without corrections due to zero-point renormalization.}
\label{table:butadiene}
\end{table}

\begin{table}[t]
\centering
  \setlength{\tabcolsep}{12pt} 
  \renewcommand{\arraystretch}{1.5} 
\begin{tabular}{cccc}
\hline
\hline
DFT level & LDA & B3LYP & HF \\
\hline
$E(\text{S}_1)_{\text{static}}$ (eV) & $5.519$ & $5.517$ & $5.66$\\
$E(\text{S}_1)_{\text{ZPR}}$ (eV) & $5.314$ & $5.341$ & $5.523$\\
$\Delta E(\text{S}_1)$ (eV) & $-0.205(30)$ & $-0.176(32)$ & $-0.137(40)$\\
\hline
\hline
\end{tabular}
\caption{Energy of the first singlet exciton of cyclopentadiene at the different levels of theory and with/without corrections due to zero-point renormalization.}
\label{table:cyclopentadiene}
\end{table}

\begin{table}[t]
\centering
  \setlength{\tabcolsep}{12pt} 
  \renewcommand{\arraystretch}{1.5} 
\begin{tabular}{cccc}
\hline
\hline
DFT level & LDA & B3LYP & HF \\
\hline
$E(\text{S}_1)_{\text{static}}$ (eV) & $6.625$ & $6.956$ & $7.243$\\
$E(\text{S}_1)_{\text{ZPR}}$ (eV) & $5.9$ & $6.141$ & $6.703$\\
$\Delta E(\text{S}_1)$ (eV) & $-0.725(31)$ & $-0.815(30)$ & $-0.540(34)$\\
\hline
\hline
\end{tabular}
\caption{Energy of the first singlet exciton of cyclopropene at the different levels of theory and with/without corrections due to zero-point renormalization.}
\label{table:cyclopropene}
\end{table}

\begin{table}[t]
\centering
  \setlength{\tabcolsep}{12pt} 
  \renewcommand{\arraystretch}{1.5} 
\begin{tabular}{cccc}
\hline
\hline
DFT level & LDA & B3LYP & HF \\
\hline
$E(\text{S}_1)_{\text{static}}$ (eV) & $3.722$  & $4.732$ & $6.096$\\
$E(\text{S}_1)_{\text{ZPR}}$ (eV) & $3.5$ & $4.471$ & $5.932$\\
$\Delta E(\text{S}_1)$ (eV) & $-0.222(76)$ & $-0.261(33)$ & $-0.164(29)$\\
\hline
\hline
\end{tabular}
\caption{Energy of the first singlet exciton of cytosine at the different levels of theory and with/without corrections due to zero-point renormalization.}
\label{table:cytosine}
\end{table}

\begin{table}[t]
\centering
  \setlength{\tabcolsep}{12pt} 
  \renewcommand{\arraystretch}{1.5} 
\begin{tabular}{cccc}
\hline
\hline
DFT level & LDA & B3LYP & HF \\
\hline
$E(\text{S}_1)_{\text{static}}$ (eV) & $9.074$ & $8.815$ & $8.392$\\
$E(\text{S}_1)_{\text{ZPR}}$ (eV) & $7.322$ & $7.712$ & $8.097$\\
$\Delta E(\text{S}_1)$ (eV) & $-1.752(31)$ & $-1.103(27)$ & $-0.295(39)$\\
\hline
\hline
\end{tabular}
\caption{Energy of the first singlet exciton of ethene at the different levels of theory and with/without corrections due to zero-point renormalization.}
\label{table:ethene}
\end{table}

\begin{table}[t]
\centering
  \setlength{\tabcolsep}{12pt} 
  \renewcommand{\arraystretch}{1.5} 
\begin{tabular}{cccc}
\hline
\hline
DFT level & LDA & B3LYP & HF \\
\hline
$E(\text{S}_1)_{\text{static}}$ (eV) & $3.824$ & $4.04$ & $4.595$\\
$E(\text{S}_1)_{\text{ZPR}}$ (eV) & $3.729$ & $3.944$ & $4.494$\\
$\Delta E(\text{S}_1)$ (eV) & $-0.095(24)$ & $-0.096(26)$ & $-0.101(25)$\\
\hline
\hline
\end{tabular}
\caption{Energy of the first singlet exciton of formaldehyde at the different levels of theory and with/without corrections due to zero-point renormalization.}
\label{table:formaldehyde}
\end{table}

\begin{table}[t]
\centering
  \setlength{\tabcolsep}{12pt} 
  \renewcommand{\arraystretch}{1.5} 
\begin{tabular}{cccc}
\hline
\hline
DFT level & LDA & B3LYP & HF \\
\hline
$E(\text{S}_1)_{\text{static}}$ (eV) & $5.57$ & $5.722$ & $6.564$\\
$E(\text{S}_1)_{\text{ZPR}}$ (eV) & $5.003$ & $5.254$ & $6.197$\\
$\Delta E(\text{S}_1)$ (eV) & $-0.567(64)$ & $-0.468(57)$ & $-0.367(45)$\\
\hline
\hline
\end{tabular}
\caption{Energy of the first singlet exciton of formamide at the different levels of theory and with/without corrections due to zero-point renormalization.}
\label{table:formamide}
\end{table}

\begin{table}[t]
\centering
  \setlength{\tabcolsep}{12pt} 
  \renewcommand{\arraystretch}{1.5} 
\begin{tabular}{cccc}
\hline
\hline
DFT level & LDA & B3LYP & HF \\
\hline
$E(\text{S}_1)_{\text{static}}$ (eV) & $6.864$ & $6.837$ & $6.787$\\
$E(\text{S}_1)_{\text{ZPR}}$ (eV) & $6.346$ & $6.51$ & $6.62$\\
$\Delta E(\text{S}_1)$ (eV) & $-0.518(25)$ & $-0.327(24)$ & $-0.167(31)$\\
\hline
\hline
\end{tabular}
\caption{Energy of the first singlet exciton of furan at the different levels of theory and with/without corrections due to zero-point renormalization.}
\label{table:furan}
\end{table}

\begin{table}[t]
\centering
  \setlength{\tabcolsep}{12pt} 
  \renewcommand{\arraystretch}{1.5} 
\begin{tabular}{cccc}
\hline
\hline
DFT level & LDA & B3LYP & HF \\
\hline
$E(\text{S}_1)_{\text{static}}$ (eV) & $5.225$ & $5.3$ & $5.572$\\
$E(\text{S}_1)_{\text{ZPR}}$ (eV) & $4.938$ & $5.128$ & $5.436$\\
$\Delta E(\text{S}_1)$ (eV) & $-0.287(19)$ & $-0.172(54)$ & $-0.136(33)$\\
\hline
\hline
\end{tabular}
\caption{Energy of the first singlet exciton of hexatriene at the different levels of theory and with/without corrections due to zero-point renormalization.}
\label{table:hexatriene}
\end{table}

\begin{table}[t]
\centering
  \setlength{\tabcolsep}{12pt} 
  \renewcommand{\arraystretch}{1.5} 
\begin{tabular}{cccc}
\hline
\hline
DFT level & LDA & B3LYP & HF \\
\hline
$E(\text{S}_1)_{\text{static}}$ (eV) & $6.629$ & $6.935$ & $7.79$\\
$E(\text{S}_1)_{\text{ZPR}}$ (eV) & $5.403$ & $5.951$ & $6.983$\\
$\Delta E(\text{S}_1)$ (eV) & $-1.226(104)$ & $-0.984(39)$ & $-0.807(32)$\\
\hline
\hline
\end{tabular}
\caption{Energy of the first singlet exciton of imidazole at the different levels of theory and with/without corrections due to zero-point renormalization.}
\label{table:imidazole}
\end{table}

\begin{table}[t]
\centering
  \setlength{\tabcolsep}{12pt} 
  \renewcommand{\arraystretch}{1.5} 
\begin{tabular}{cccc}
\hline
\hline
DFT level & LDA & B3LYP & HF \\
\hline
$E(\text{S}_1)_{\text{static}}$ (eV) & $4.751$ & $5.174$ & $5.939$\\
$E(\text{S}_1)_{\text{ZPR}}$ (eV) & $4.615$ & $5.04$ & $5.799$\\
$\Delta E(\text{S}_1)$ (eV) & $-0.136(27)$ & $-0.134(30)$ & $-0.140(36)$\\
\hline
\hline
\end{tabular}
\caption{Energy of the first singlet exciton of norbornadiene at the different levels of theory and with/without corrections due to zero-point renormalization.}
\label{table:norbornadiene}
\end{table}

\begin{table}[t]
\centering
  \setlength{\tabcolsep}{12pt} 
  \renewcommand{\arraystretch}{1.5} 
\begin{tabular}{cccc}
\hline
\hline
DFT level & LDA & B3LYP & HF \\
\hline
$E(\text{S}_1)_{\text{static}}$ (eV) & $4.454$ & $4.54$ & $4.879$\\
$E(\text{S}_1)_{\text{ZPR}}$ (eV) & $4.039$ & $4.451$ & $4.778$\\
$\Delta E(\text{S}_1)$ (eV) & $-0.415(73)$ & $-0.089(15)$ & $-0.101(35)$\\
\hline
\hline
\end{tabular}
\caption{Energy of the first singlet exciton of octatetraene at the different levels of theory and with/without corrections due to zero-point renormalization.}
\label{table:octatetraene}
\end{table}

\begin{table}[t]
\centering
  \setlength{\tabcolsep}{12pt} 
  \renewcommand{\arraystretch}{1.5} 
\begin{tabular}{cccc}
\hline
\hline
DFT level & LDA & B3LYP & HF \\
\hline
$E(\text{S}_1)_{\text{static}}$ (eV) & $3.521$ & $4.046$ & $6.118$\\
$E(\text{S}_1)_{\text{ZPR}}$ (eV) & $3.343$ & $3.891$ & $5.015$\\
$\Delta E(\text{S}_1)$ (eV) & $-0.178(16)$ & $-0.155(14)$ & $-1.103(13)$\\
\hline
\hline
\end{tabular}
\caption{Energy of the first singlet exciton of pyrazine at the different levels of theory and with/without corrections due to zero-point renormalization.}
\label{table:pyrazine}
\end{table}

\begin{table}[t]
\centering
  \setlength{\tabcolsep}{12pt} 
  \renewcommand{\arraystretch}{1.5} 
\begin{tabular}{cccc}
\hline
\hline
DFT level & LDA & B3LYP & HF \\
\hline
$E(\text{S}_1)_{\text{static}}$ (eV) & $4.309$ & $4.887$ & $6.275$\\
$E(\text{S}_1)_{\text{ZPR}}$ (eV) & $4.038$ & $4.669$ & $5.924$\\
$\Delta E(\text{S}_1)$ (eV) & $-0.271(22)$ & $-0.218(22)$ & $-0.351(22)$\\
\hline
\hline
\end{tabular}
\caption{Energy of the first singlet exciton of pyridine at the different levels of theory and with/without corrections due to zero-point renormalization.}
\label{table:pyridine}
\end{table}

\begin{table}[t]
\centering
  \setlength{\tabcolsep}{12pt} 
  \renewcommand{\arraystretch}{1.5} 
\begin{tabular}{cccc}
\hline
\hline
DFT level & LDA & B3LYP & HF \\
\hline
$E(\text{S}_1)_{\text{static}}$ (eV) & $3.727$ & $4.335$ & $6.589$\\
$E(\text{S}_1)_{\text{ZPR}}$ (eV) & $3.511$ & $4.13$ & $5.72$\\
$\Delta E(\text{S}_1)$ (eV) & $-0.216(23)$ & $-0.205(25)$ & $-0.869(27)$\\
\hline
\hline
\end{tabular}
\caption{Energy of the first singlet exciton of pyrimidine at the different levels of theory and with/without corrections due to zero-point renormalization.}
\label{table:pyrimidine}
\end{table}

\begin{table}[t]
\centering
  \setlength{\tabcolsep}{12pt} 
  \renewcommand{\arraystretch}{1.5} 
\begin{tabular}{cccc}
\hline
\hline
DFT level & LDA & B3LYP & HF \\
\hline
$E(\text{S}_1)_{\text{static}}$ (eV) & $7.049$ & $7.03$ & $7.334$\\
$E(\text{S}_1)_{\text{ZPR}}$ (eV) & $4.89$ & $5.668$ & $6.678$\\
$\Delta E(\text{S}_1)$ (eV) & $-2.159(206)$ & $-1.362(43)$ & $-0.656(31)$\\
\hline
\hline
\end{tabular}
\caption{Energy of the first singlet exciton of pyrrole at the different levels of theory and with/without corrections due to zero-point renormalization.}
\label{table:pyrrole}
\end{table}

\begin{table}[t]
\centering
  \setlength{\tabcolsep}{12pt} 
  \renewcommand{\arraystretch}{1.5} 
\begin{tabular}{cccc}
\hline
\hline
DFT level & LDA & B3LYP & HF \\
\hline
$E(\text{S}_1)_{\text{static}}$ (eV) & $1.747$ & $2.303$ & $3.507$\\
$E(\text{S}_1)_{\text{ZPR}}$ (eV) & $1.697$ & $2.242$ & $3.423$\\
$\Delta E(\text{S}_1)$ (eV) & $-0.050(14)$ & $-0.061(13)$ & $-0.084(13)$\\
\hline
\hline
\end{tabular}
\caption{Energy of the first singlet exciton of tetrazine at the different levels of theory and with/without corrections due to zero-point renormalization.}
\label{table:tetrazine}
\end{table}

\begin{table}[t]
\centering
  \setlength{\tabcolsep}{12pt} 
  \renewcommand{\arraystretch}{1.5} 
\begin{tabular}{cccc}
\hline
\hline
DFT level & LDA & B3LYP & HF \\
\hline
$E(\text{S}_1)_{\text{static}}$ (eV) & $4.831$ & $5.334$ & $6.454$\\
$E(\text{S}_1)_{\text{ZPR}}$ (eV) & $3.995$ & $4.626$ & $6.1$\\
$\Delta E(\text{S}_1)$ (eV) & $-0.836(42)$ & $-0.708(32)$ & $-0.354(27)$\\
\hline
\hline
\end{tabular}
\caption{Energy of the first singlet exciton of thymine at the different levels of theory and with/without corrections due to zero-point renormalization.}
\label{table:thymine}
\end{table}

\begin{table}[t]
\centering
  \setlength{\tabcolsep}{12pt} 
  \renewcommand{\arraystretch}{1.5} 
\begin{tabular}{cccc}
\hline
\hline
DFT level & LDA & B3LYP & HF \\
\hline
$E(\text{S}_1)_{\text{static}}$ (eV) & $3.697$ & $4.476$ & $7.589$\\
$E(\text{S}_1)_{\text{ZPR}}$ (eV) & $3.402$ & $4.091$ & $5.923$\\
$\Delta E(\text{S}_1)$ (eV) & $-0.295(22)$ & $-0.385(24)$ & $-1.666(27)$\\
\hline
\hline
\end{tabular}
\caption{Energy of the first singlet exciton of triazine at the different levels of theory and with/without corrections due to zero-point renormalization.}
\label{table:triazine}
\end{table}

\begin{table}[t]
\centering
  \setlength{\tabcolsep}{12pt} 
  \renewcommand{\arraystretch}{1.5} 
\begin{tabular}{cccc}
\hline
\hline
DFT level & LDA & B3LYP & HF \\
\hline
$E(\text{S}_1)_{\text{static}}$ (eV) & $3.919$ & $5.477$ & $6.589$\\
$E(\text{S}_1)_{\text{ZPR}}$ (eV) & $3.749$  & $4.5$ & $6.109$\\
$\Delta E(\text{S}_1)$ (eV) & $-0.170(30)$ & $-0.977(29)$ & $-0.480(22)$\\
\hline
\hline
\end{tabular}
\caption{Energy of the first singlet exciton of uracil at the different levels of theory and with/without corrections due to zero-point renormalization.}
\label{table:uracil}
\end{table}

\begin{table}[t]
\centering
  \setlength{\tabcolsep}{12pt} 
  \renewcommand{\arraystretch}{1.5} 
\begin{tabular}{cccc}
\hline
\hline
DFT level & LDA & B3LYP & HF \\
\hline
$E(\text{S}_1)_{\text{static}}$ (eV) & $5.361$ & $5.521$ & $6.197$\\
$E(\text{S}_1)_{\text{ZPR}}$ (eV) & $5.038$ & $5.408$ & $6.12$\\
$\Delta E(\text{S}_1)$ (eV) & $-0.323(184)$ & $-0.113(14)$ & $-0.077(14)$\\
\hline
\hline
\end{tabular}
\caption{Energy of the first singlet exciton of benzene at the different levels of theory and with/without corrections due to zero-point renormalization.}
\label{table:benzene}
\end{table}

\begin{table}[t]
\centering
  \setlength{\tabcolsep}{12pt} 
  \renewcommand{\arraystretch}{1.5} 
\begin{tabular}{cccc}
\hline
\hline
DFT level & LDA & B3LYP & HF \\
\hline
$E(\text{S}_1)_{\text{static}}$ (eV) & $4.171$ & $4.421$ & $4.917$\\
$E(\text{S}_1)_{\text{ZPR}}$ (eV) & $4.047$  & $4.286$ & $4.799$\\
$\Delta E(\text{S}_1)$ (eV) & $-0.124(12)$ & $-0.135(20)$ & $-0.118(22)$\\
\hline
\hline
\end{tabular}
\caption{Energy of the first singlet exciton of naphthalene at the different levels of theory and with/without corrections due to zero-point renormalization.}
\label{table:naphthalene}
\end{table}

\begin{table}[t]
\centering
  \setlength{\tabcolsep}{12pt} 
  \renewcommand{\arraystretch}{1.5} 
\begin{tabular}{cccc}
\hline
\hline
DFT level & LDA & B3LYP & HF \\
\hline
$E(\text{S}_1)_{\text{static}}$ (eV) & $3.145$ & $3.449$ & $4.131$\\
$E(\text{S}_1)_{\text{ZPR}}$ (eV) & $3.084$ & $3.384$ & $4.035$\\
$\Delta E(\text{S}_1)$ (eV) & $-0.061(15)$ & $-0.065(18)$ & $-0.096(26)$\\
\hline
\hline
\end{tabular}
\caption{Energy of the first singlet exciton of anthracene at the different levels of theory and with/without corrections due to zero-point renormalization.}
\label{table:anthracene}
\end{table}

\begin{table}[t]
\centering
  \setlength{\tabcolsep}{12pt} 
  \renewcommand{\arraystretch}{1.5} 
\begin{tabular}{cccc}
\hline
\hline
DFT level & LDA & B3LYP & HF \\
\hline
$E(\text{S}_1)_{\text{static}}$ (eV) & $2.39$ & $2.687$ & $3.375$\\
$E(\text{S}_1)_{\text{ZPR}}$ (eV) & $2.346$ & $2.643$ & $3.284$\\
$\Delta E(\text{S}_1)$ (eV) & $-0.044(15)$ & $-0.044(12)$ & $-0.091(30)$\\
\hline
\hline
\end{tabular}
\caption{Energy of the first singlet exciton of tetracene at the different levels of theory and with/without corrections due to zero-point renormalization.}
\label{table:tetracene}
\end{table}

\begin{table}[t]
\centering
  \setlength{\tabcolsep}{12pt} 
  \renewcommand{\arraystretch}{1.5} 
\begin{tabular}{cccc}
\hline
\hline
DFT level & LDA & B3LYP & HF \\
\hline
$E(\text{S}_1)_{\text{static}}$ (eV) & $1.859$ & $2.154$ & $2.812$\\
$E(\text{S}_1)_{\text{ZPR}}$ (eV) & $1.827$ & $2.122$ & $2.741$\\
$\Delta E(\text{S}_1)$ (eV) & $-0.032(13)$ & $-0.032(16)$ & $-0.071(29)$\\
\hline
\hline
\end{tabular}
\caption{Energy of the first singlet exciton of pentacene at the different levels of theory and with/without corrections due to zero-point renormalization.}
\label{table:pentacene}
\end{table}

\begin{figure}[tb]
\centering
\includegraphics[width=0.9\linewidth]{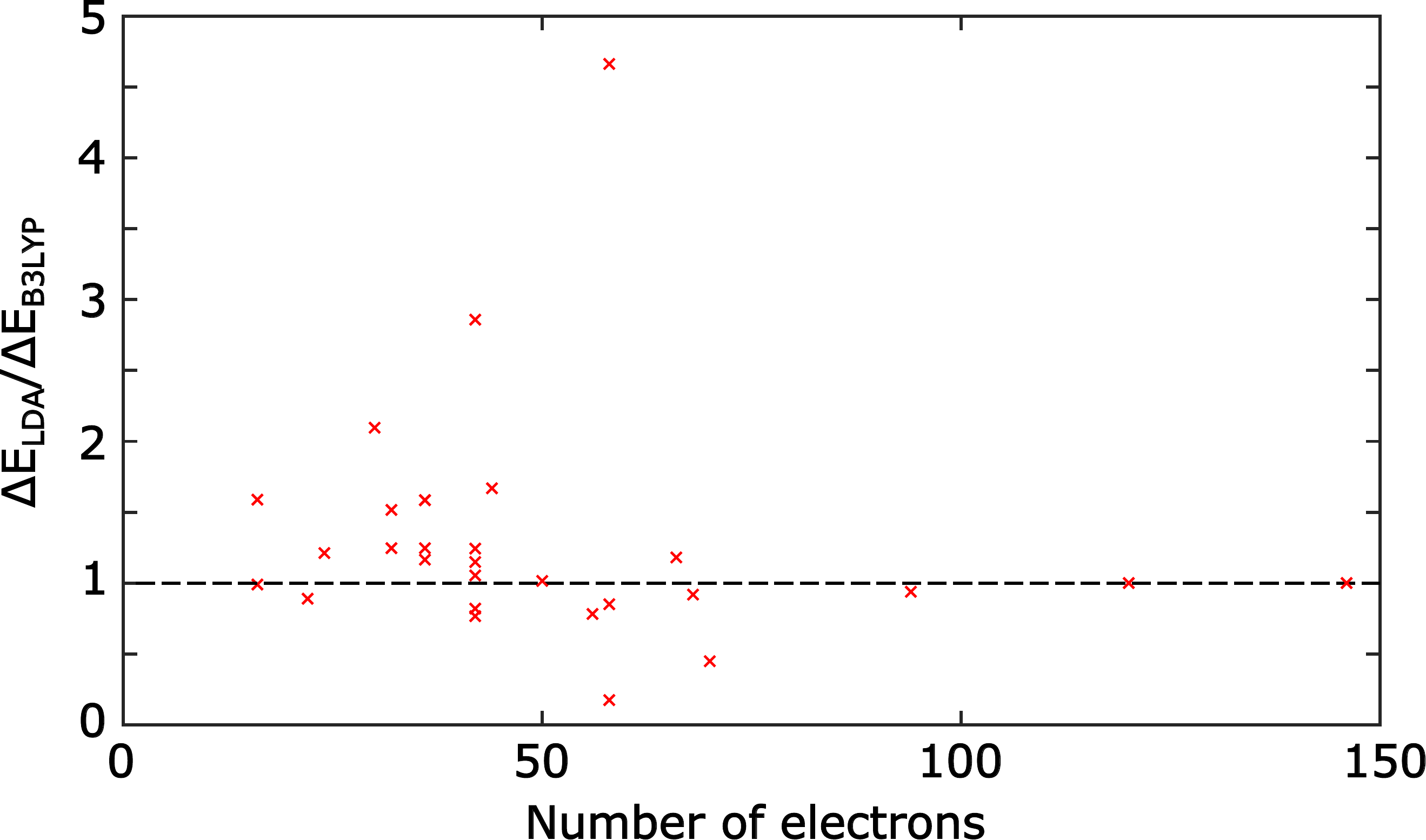}
\caption{Ratio of LDA and B3LYP zero-point renormalization to the excited state energy of the different molecules, as a function of number of electrons in the studied system.}
\label{fig:LDA_vs_B3LYP}
\end{figure}

Figure\,\ref{fig:LDA_vs_B3LYP} visualizes the ratio of the LDA to the B3LYP correction to the excited state energies due to zero-point vibrational motion.

\clearpage

\section{Comparison of Monte Carlo energies to experiment}

\begin{table}[t]
\centering
  \setlength{\tabcolsep}{12pt} 
  \renewcommand{\arraystretch}{1.5} 
\begin{tabular}{ccc}
\hline
\hline
molecule & $E(\text{S}_1)$ (eV) & Reference \\
\hline
acetamide & $5.44$ & \cite{Nielsen1967} \\
acetone & $4.38$ & \cite{Walzl1987} \\
adenine & $4.63$ & \cite{Voelter1968} \\
benzene & $4.9$ & \cite{Atsunari1991} \\
benzoquinone & $2.7$ & \cite{Weber2001} \\
butadiene & $5.92$ & \cite{Leopold1984}\\
cyclopentadiene & $5.3$ & \cite{Sabljic1990} \\
cyclopropene & $6.45$ & \cite{Robin1969} \\
cytosine & $4.6$ & \cite{Raksanyl1978} \\
ethene & $7.66$ & \cite{Mulliken1977} \\
formaldehyde & $3.79$ & \cite{Walzl1987} \\
formamide & $5.5$ & \cite{Clark1995} \\
furan & $6.06$ & \cite{Flicker1976} \\
hexatriene & $4.93$ & \cite{Leopold1984} \\
naphthalene & $3.97$ & \cite{George1968} \\
norbornadiene & $5.23$ & \cite{Frueholz1979} \\
octatetraene & $4.41$ & \cite{Gavin1978} \\
pyrazine & $3.83$ & \cite{Bolovinos1984} \\
pyridine & $4.59$ & \cite{Bolovinos1984} \\
pyrimidine & $4.16$ & \cite{Bolovinos1984} \\
pyrrole & $5.98$ & \cite{Flicker1976} \\
tetrazine & $2.34$ & \cite{Mason1959} \\
thymine & $4.9-5.2$ & \cite{Novros1986} \\
uracil & $4.58$ & \cite{Brady1988} \\
\hline
\hline
\end{tabular}
\caption{Experimental values for the first excited singlet state of
the studied molecules (those for which we were able to find experimental
data for this state), and references to the relevant experimental studies.}
\label{table:experiment}
\end{table}

Table\,\ref{table:experiment} summarizes the experimental values for the
first single excitation $E(\text{S}_1)$ of the studied molecules, as these are also summarized in Ref. \cite{Schreiber2008}. The relevant references to the
experimental studies are also given.
Figure\,3 of the main manuscript provides a visualization of
computed vs experimental values of the exciton energy, for those
molecules for which we found experimental data. 
This is done at three levels of theory: static TD-DFT (B3LYP),
static wavefunction-based methods as obtained from the benchmark
values on the Thiel set \cite{Schreiber2008}, and TD-DFT
including the effects of zero-point renormalization due to nuclear quantum motion (B3LYP+ZPR). 
The closer a point lies to the $y=x$ line, the better the agreement with experiment. As a means of visualizing the accuracy
of the three different methods we fit a linear model $y=ax+b$
for the three cases. The $a$ and $b$ parameters of these linear
fits are given for the three different theory levels in Table\,\ref{table:fit_parameters}, along with the correlation
coefficient $r^2$. Clearly, the B3LYP+ZPR fit
is the closest to the $y=x$ line, since the slope $a$ is the closest to the value of one and the offset $b$ is $66$\,meV (compared to offsets of $316$\,meV and $639$\,meV for Thiel and static B3LYP values respectively). In all three cases,
a linear fit provides a good description  of the scatter plots,
as the correlation coefficient $r^2$ assumes values between $0.96$ and $0.98$. Table\,\ref{table:stdev} gives the standard
deviation to the average values of the statistical measures
given in Table\,1 of the main manuscript.

\begin{table}[t]
\centering
  \setlength{\tabcolsep}{12pt} 
  \renewcommand{\arraystretch}{1.5} 
\begin{tabular}{cccc}
\hline
\hline
Theory level & Thiel & B3LYP & B3LYP+ZPR \\
\hline
$a$ & $1.124$ & $1.207$ & $1.016$ \\
$b$ (eV) & $-0.316$ & $-0.639$ & $-0.066$ \\
$r^2$ & $0.98$ & $0.97$ & $0.96$ \\
\hline
\hline
\end{tabular}
\caption{Parameters for the linear model fits $y=ax+b$ appearing in Figure\,3 of the main manuscript.}
\label{table:fit_parameters}
\end{table}

\begin{table}[t]
\centering
  \setlength{\tabcolsep}{8pt} 
\begin{tabular}{ccccc}
\hline
\hline
& $\sigma(\text{bias})$ (eV) & $\sigma(\text{rel.bias})$ & $\sigma(\text{RMSE})$ (eV) & $\sigma(\text{rel.RMSE})$ \\
\hline
B3LYP & $0.342$ & $0.062$ & $0.601$ & $0.100$\\
Thiel & $0.223$ & $0.040$ & $0.440$ & $0.068$\\
B3LYP+ZPR & $0.235$ & $0.051$ & $0.260$ & $0.059$\\
\hline
\hline
\end{tabular}
\caption{Standard deviation around the average values of the (relative) bias and the (relative) root mean-squared error (RMSE) that are given in Table\,1 of the main manuscript.}
\label{table:stdev}
\end{table}

\clearpage

\section{Effect of temperature on exciton energies}

\begin{figure}[tb]
\centering
\includegraphics[width=0.7\linewidth]{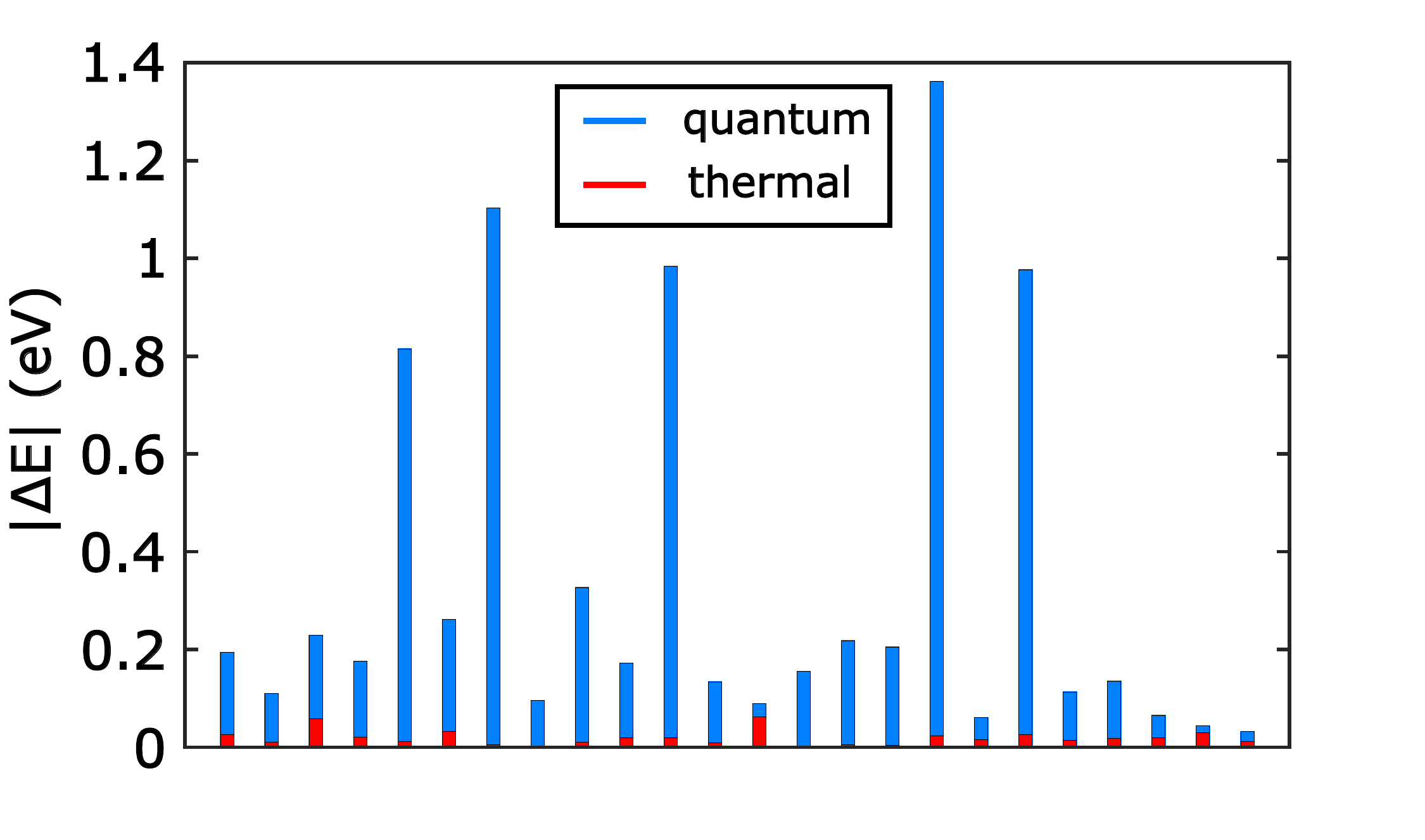}
\caption{Absolute shift of the exciton energy due to quantum (blue) and thermal (red) fluctuations.}
\label{fig:quantum_vs_thermal}
\end{figure}

Let us compare the renormalization of the energy of the first
excited state of the studied molecules due to molecular vibrations, as
computed using Monte Carlo sampling at $0$\,K and $300$\,K. In Figure\,\ref{fig:quantum_vs_thermal} we plot the absolute renormalization $\Delta E_{\text{quantum}}=|\Delta E(0\hspace{0.1cm}\text{K})|$ (blue) due to the zero-point
motion of the nuclei, versus the additional renormalization $\Delta E_{\text{thermal}}=|\Delta E(300\hspace{0.1cm}\text{K})-\Delta E(0\hspace{0.1cm}\text{K})|$ (red) that is present at $300$\,K due to the thermal activation
of vibrational modes. Every bar in Figure\,\ref{fig:quantum_vs_thermal} represents a different molecule. The first observation is that the correction to the static
exciton energy due to quantum nuclear fluctuations is in almost every
case significantly larger than the one that arises from thermal motion. This is not surprising
given the typical vibrational mode frequencies of small organic molecules (such as the ones studied here), which
for most structures lie well above the threshold of activation at room
temperature, due to the light mass of carbon and hydrogen atoms. 
It is only for the more bulky molecules from the studied structures that
the quantum and thermal corrections are comparable, as in the case of 
octatetraene with values of $\Delta E_{\text{quantum}}=89$\,meV and $\Delta
E_{\text{thermal}}=$\,62 meV. 
Since ZPR dominates
the exciton energy correction due to vibrations, we neglect thermal effects
when comparing our computed exciton energies to
experiment and previous computational studies.

\clearpage

\section{Convergence of the Monte Carlo calculations}

\begin{figure}[tb]
\centering
\includegraphics[width=\linewidth]{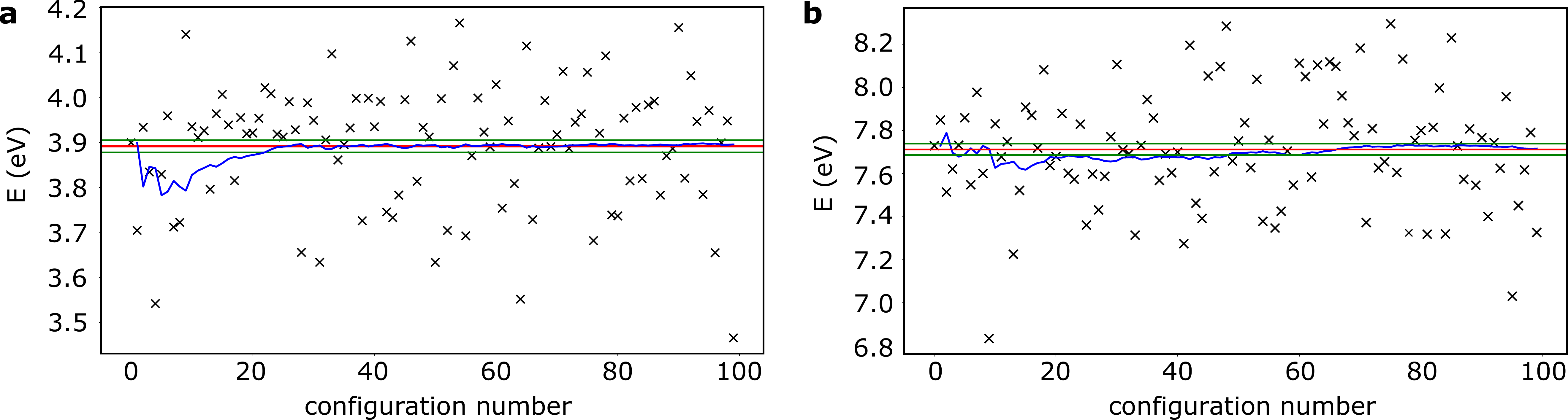}
\caption{Convergence of the Monte Carlo sampling of the exciton energy for pyrazine (panel \textbf{a}) and ethene (panel \textbf{b}). The average exciton energy is denoted in red, and its statistical uncertainty boundaries in green. We also include the running average (blue), which visualizes the convergence
of the average to its final value with increasing number of configurations.}
\label{fig:MC_convergence}
\end{figure}

The convergence of the Monte Carlo simulations for the cases of pyrazine
and ethene is visualized in Figures\,\ref{fig:MC_convergence}a and 
\ref{fig:MC_convergence}b respectively. The blue line denotes the running 
average, which converges to the final average value of the exciton energy 
(red) as we sample more configurations. The statistical uncertainty of this
sampling technique is small, and its boundaries are highlighted with
green lines. As for the cases of pyrazine and ethene visualized here,
for the vast majority of structures the Monte Carlo sampling converges
when we include approximately $50$ configurations.

\clearpage

\section{Mode-resolved quadratic correction to the exciton energies}

\begin{table}[t]
\centering
  \setlength{\tabcolsep}{8pt} 
\begin{tabular}{cccc}
\hline
\hline
molecule & $\Delta E(\text{S}_1)_{\text{quadratic}}$ (eV) & $\omega_{\text{dominant}}$ ($\text{cm}^{-1}$) & contribution of dominant mode (\%) \\
\hline
acetamide & $-0.216$ & $36$ & $29$ \\
acetone & $-0.104$ & $493$ & $18$ \\
adenine & $-0.192$ & $1652$ & $13$ \\
anthracene & $-0.061$ & $391$ & $8$ \\
benzene & $-0.105$ & $723$ & $17$ \\
benzoquinone & $-0.115$ & $1753$ & $20$ \\
butadiene & $-0.203$ & $783$ & $24$ \\
cyclopentadiene & $-0.146$ & $3059$ & $21$ \\
cyclopropene & $-0.824$ & $863$ & $37$ \\
cytosine & $-0.281$ & $1813$ & $29$ \\
ethene & $-1.307$ & $1064$ & $59$ \\
formaldehyde & $-0.084$ & $1193$ & $54$ \\
formamide & $-0.353$ & $67$ & $73$ \\
furan & $-0.317$ & $1208$ & $22$ \\
hexatriene & $-0.119$ & $713$ & $15$ \\
imidazole & $-1.018$ & $520$ & $15$ \\
naphthalene & $-0.101$ & $513$ & $12$ \\
norbornadiene & $-0.145$ & $908$ & $10$ \\
octatetraene & $-0.086$ & $685$ & $11$ \\
pentacene & $-0.029$ & $476$ & $7$ \\
pyrazine & $-0.154$ & $1588$ & $35$ \\
pyridine & $-0.218$ & $1633$ & $38$ \\
pyrimidine & $-0.206$ & $1626$ & $39$ \\
pyrrole & $-1.419$ & $476$ & $10$ \\
tetracene & $-0.044$ & $386$ & $6$ \\
tetrazine & $-0.065$ & $1574$ & $62$ \\
thymine & $-0.702$ & $913$ & $8$ \\
triazine & $-0.372$ & $1609$ & $30$ \\
uracil & $-1.030$ & $977$ & $9$ \\
\hline
\hline
\end{tabular}
\caption{Quadratic correction to the energy of the first singlet exciton
of the studied systems. For each molecule, the frequency $\omega_{\text{dominant}}$ of the mode which most strongly renormalizes the exciton energy, along with the percentage
of the total renormalization that it causes, is given.}
\label{table:quadratic_correction}
\end{table}

In Table\,\ref{table:quadratic_correction} the correction to the first
exciton energy of the studied molecules is given, as computed within the
quadratic approximation. The quadratic approximation allows for a mode-resolved picture of this effect, hence we also give the frequency
of the mode with the largest contribution to the exciton energy renormalization, along with the percentage of the renormalization for
which it is responsible. 

\clearpage 

\bibliographystyle{unsrt}
\bibliography{references}